\documentclass[aps,prx,preprint,onecolumn,citeautoscript,footinbib,superscriptaddress,
eqsecnum]{revtex4-1}
\usepackage{amsmath,amssymb}
\usepackage{graphicx}% Include figure files
\usepackage{dcolumn}% Align table columns on decimal point
\usepackage{bm}% bold math
\usepackage{braket}
\usepackage{verbatim}
\usepackage{float}
\usepackage[caption = false]{subfig}
\usepackage{color}
\usepackage{tcolorbox}
\usepackage{xcolor}
\usepackage{relsize}
\usepackage{amsthm}
\usepackage{enumerate}
\usepackage{physics}
\usepackage{soul,xcolor}
\usepackage[T1]{fontenc}

\newcommand{\beq}{\begin{eqnarray}}
\newcommand{\eeq}{\end{eqnarray}}

\newcommand{\bsp}{\begin{split}}
\newcommand{\esp}{\end{split}}

\newcommand{\be}{\begin{equation}}
\newcommand{\ee}{\end{equation}}

\newcommand{\js}{J_{s}}
\newcommand{\jp}{J_{\perp}}
\newcommand{\jc}{J_{3}}

\newcommand{\ec}[1]{\epsilon_{c #1}}
\newcommand{\ef}[1]{\epsilon_{f #1}}
\newcommand{\ez}[1]{\epsilon_{Z #1}}

\newcommand{\Ea}[1]{E_{1 #1}}
\newcommand{\Eb}[1]{E_{2 #1}}
\newcommand{\Ez}[1]{E_{Z #1}}

\newcommand{\cd}[1]{c^{\dagger}_{#1}}
\newcommand{\fd}[1]{f^{\dagger}_{#1}}
\newcommand{\zd}[1]{Z^{*}_{#1}}
\newcommand{\gc}[2]{G_{c}(#1,#2)}
\newcommand{\gf}[2]{G_{f}(#1,#2)}
\newcommand{\gz}[2]{D(#1,#2)}
\newcommand{\gzzero}[2]{D^{0}(#1,#2)}
\newcommand{\gcf}[2]{G_{cf}(#1,#2)}
\newcommand{\gfc}[2]{G_{fc}(#1,#2)}

\newcommand{\gcc}[2]{\frac{#2 - \ef{#1}}{(#2 -\Ea{#1})(#2 - \Eb{#1})}}
\newcommand{\gff}[2]{\frac{#2 - \ec{#1}}{(#2 -\Ea{#1})(#2 - \Eb{#1})}}
\newcommand{\gzf}[2]{\frac{g}{(#2 +\Ez{#1})(-#2 + \Ez{#1})}}

\newcommand{\nf}[1]{n_{F}(#1)}
\newcommand{\nb}[1]{n_{B}(#1)}

\newcommand{\kk}{\vec{k}}
\newcommand{\qp}{\vec{Q}_{\pi}}

\newcommand{\iw}{i\omega}

\makeatletter
\newcommand*{\rom}[1]{\expandafter\@slowromancap\romannumeral #1@}
\makeatother

\graphicspath{{./images/}}
\def\bea{\begin{eqnarray}}
\def\eea{\end{eqnarray}}

\usepackage{mathtools} % for cases
\makeatletter

\def\env@sqcases{%
  \let\@ifnextchar\new@ifnextchar
  \left\lbrack
  \def\arraystretch{1.2}%
  \array{@{}l@{\quad}l@{}}%
}
\makeatother

\usepackage{color}
\usepackage[papersize={8.5in,11in}]{geometry}
\usepackage[colorlinks=true]{hyperref}
\hypersetup{
    bookmarks=true,         % show bookmarks bar?
    unicode=false,          % non-Latin characters
    pdftoolbar=true,        % show Acrobat
    pdfmenubar=true,        % show Acrobat
    pdffitwindow=false,     % window fit to page when opened
    pdfstartview={FitH},    % fits the width of the page to the window
    pdftitle={S},    % title
    pdfauthor={S. Sachdev},     % author
    pdfsubject={},   % subject of the document
    pdfcreator={},   % creator of the document
    pdfproducer={}, % producer of the document
    pdfkeywords={} {} {}, % list of keywords
    pdfnewwindow=true,      % links in new window
    colorlinks=true,       % false: boxed links; true: colored links
    linkcolor=magenta, %red,          % color of internal links (change box color with linkbordercolor)
    citecolor=blue,        % color of links to bibliography
    filecolor=magenta,      % color of file links
    urlcolor=blue           % color of external links
}

\begin{document}

\setstcolor{red}

\title{Spin density wave, Fermi liquid, and fractionalized phases\\ in a theory of antiferromagnetic metals\\ using paramagnons and bosonic spinons}
\author{Alexander Nikolaenko}
\affiliation{Department of Physics, Harvard University, Cambridge, MA-02138, USA}
\author{Jonas von Milczewski}
\affiliation{Department of Physics, Harvard University, Cambridge, MA-02138, USA}
\affiliation{Max-Planck-Institute of Quantum Optics, 85748 Garching, Germany}
\affiliation{Munich Center for Quantum Science and Technology, 80799 Munich, Germany}
\author{Darshan G. Joshi}
\affiliation{Department of Physics, Harvard University, Cambridge, MA-02138, USA}
\affiliation{Tata Institute of Fundamental Research, Hyderabad 500046, India}
\author{Subir Sachdev}
\affiliation{Department of Physics, Harvard University, Cambridge, MA-02138, USA}
%\affiliation{School of Natural Sciences, Institute for Advanced Study, Princeton, NJ-08540, USA}

\date{\today.~
\href{https://arxiv.org/abs/2211.10452}{arXiv:2211.10452}}
% It is always \today, today,
             %  but any date may be explicitly specified

\begin{abstract}
The pseudogap metal phase of the hole-doped cuprates can be described by small Fermi surfaces of electron-like quasiparticles, which enclose a volume violating the Luttinger relation. This violation requires the existence of additional fractionalized excitations which can be viewed as fractionalized remnants of the paramagnon. 
We fractionalize the paramagnon into the bosonic spinons of the spin liquid described by the $\mathbb{CP}^1$ U(1) gauge theory,
and present a gauge theory of the bosonic spinons, a Higgs field, and an ancilla layer of fermions coupled to the original electrons.  Along with the small Fermi surface pseudogap metal, this theory displays conventional phases: the large Fermi surface Fermi liquid with a low-energy paramagnon mode, and phases with spin density wave order. We describe the evolution of the electronic photoemission spectrum across these quantum phase transitions. We consider both the two-sublattice N\'eel and incommensurate spin density wave phases, and find that the latter has spiral spin correlations.
\end{abstract}

\pacs{Valid PACS appear here}% PACS, the Physics and Astronomy
                             % Classification Scheme.
%\keywords{Suggested keywords}%Use showkeys class option if keyword
                              %display desired
\maketitle{}

\section{Introduction}

Paramagnons are central actors in the theory of magnetism in Fermi liquids \cite{Berk66,Doniach66}: they are collective spin excitations with spin $S=1$, Landau-damped by their coupling to gapless particle-hole excitations across the Fermi surface. Exchange of ferromagnetic paramagnons is believed to lead to superfluidity in $^3$He \cite{vollhardt}, and exchange of antiferromagnetic paramagnons is argued to lead to superconductivity with unconventional spin-singlet pairing in numerous correlated electron compounds \cite{Scalapino95,Chubukov11}. 

The application of the paramagnon theory to the hole-doped cuprates 
faces a challenge from the experimentally observed pseudogap metal regime at low hole doping away from the insulating antiferromagnet at half-filling. This is a metallic phase with no long-range magnetic order in which the conventional Luttinger-volume Fermi surface is partially gapped, displaying only `Fermi arcs' in the photoemission spectrum \cite{ShenShen,He2011,Hoffman14,Davis14,chen2019incoherent}. In many theoretical approaches, including the one followed in the present paper, 
the pseudogap metal is postulated to have small `hole pocket' Fermi surfaces of size $p$, where $p$ is the hole-doping density (there is photoemission evidence for such pockets \cite{PDJ11}). Such a pseudogap metal appeared early on in Ref.~\cite{SS93}, in a theory of fluctuating paramagnons in a doped antiferromagnet. 
This metallic state, if continued to zero temperature ($T$), does not obey the Luttinger theorem on the volume enclosed by the Fermi surface, which states that the Fermi surface of holes must have size $1+p$ (or a Fermi surface of electrons must have size $1-p$).
It was argued \cite{FLS1,FLS2} that violations of the Luttinger theorem in such a metal, hereafter called FL* in this paper, require the presence of fractionalization and emergent gauge fields: in particular, any such metallic state must have deconfined, charge 0, spin $S=1/2$ excitations (`spinons'). These spinons are distinct from the quasiparticle excitations around the Fermi surface of holes, which have charge $+e$ and spin $1/2$. The theory of Ref.~\cite{SS93} was extended to a complete theory of the FL* metal in Refs.~\cite{QS10,MS11,Punk12,SS18a} by fractionalizing the O(3) paramagnon field ${\bm n} \equiv n^a$ ($a =x,y,z$ is a spin index) in a $\mathbb{CP}^1$ representation \cite{rs1,rs2} by
\begin{equation}
    {\bm n} =w^*_\alpha \, {\bm \sigma}^\alpha_{~\beta} \, w^\beta \,. \label{i1}
\end{equation}
Here $w^\beta$ are the required bosonic spinons, with $\alpha, \beta = \uparrow, \downarrow$ $S=1/2$ spin indices, and ${\bm \sigma}$ are the Pauli matrices (we also note a theory of the FL* state using fermionic spinons \cite{RW06}). Note that (\ref{i1}) introduces a U(1) gauge invariance, and the resulting U(1) photon is the emergent photon of the FL* metal. The monopoles in this U(1) gauge field carry Berry phases \cite{rs1,rs2}, and this extends the range of deonfinement \cite{senthil1,senthil2}.

\begin{figure}
\begin{center}
\includegraphics[width=6.5in]{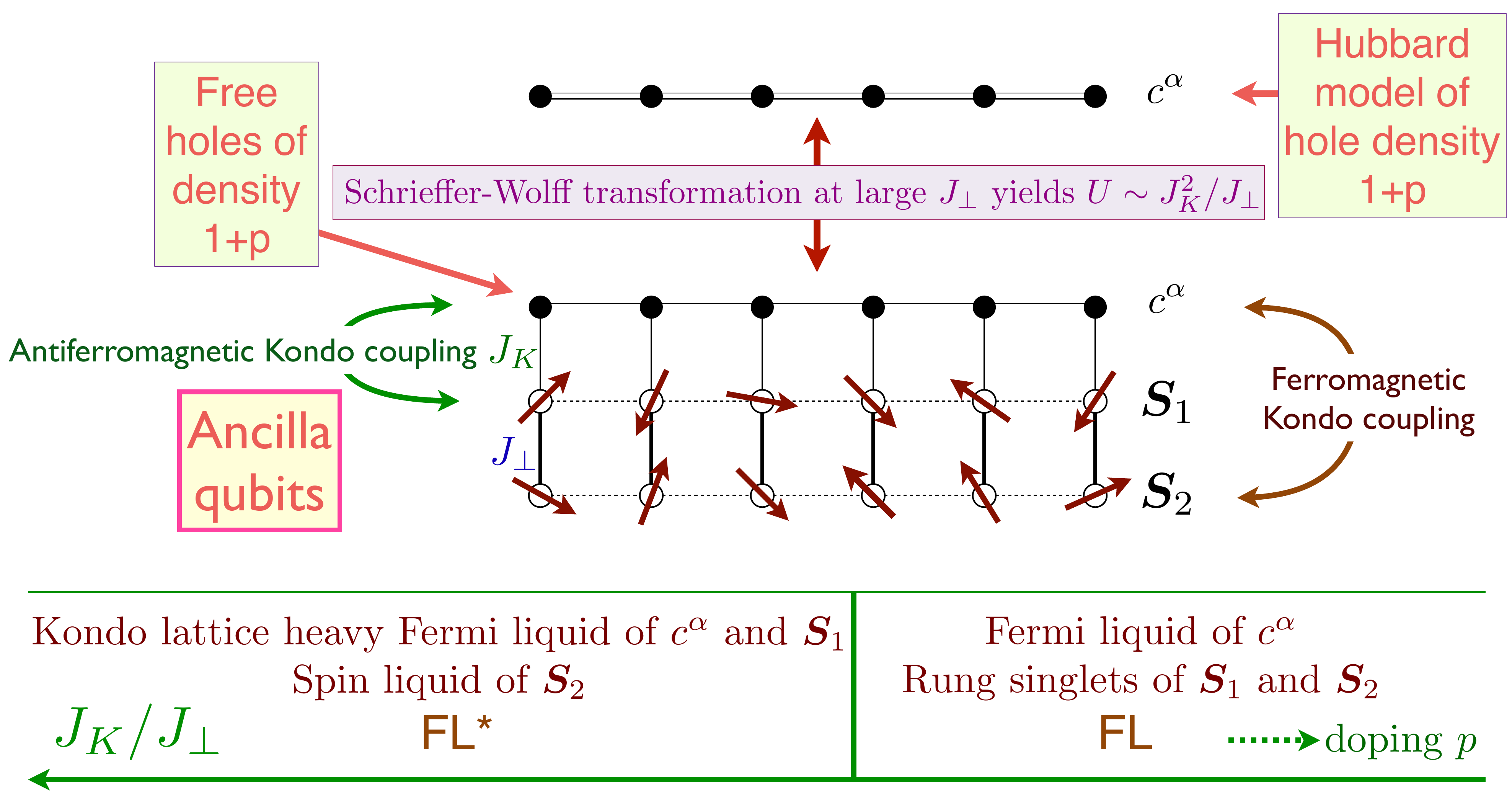}
\end{center}
\caption{Schematic illustration of the ancilla theory of the single band Hubbard model \cite{Zhang2020,Zhang2021,nikolaenko2021,Mascot22}. A canonical transformation, which can be carried out order-by-order in $J_\perp$, maps free electrons ($c^\alpha$) coupled to a bilayer antiferromagnet (with $S=1/2$ spins ${\bm S}_1$,${\bm S}_2$ in the two layers respectively) to a Hubbard model for $c^\alpha$ with on-site repulsion $U$. In the present paper we employ fermionic spinons to describe the ${\bm S}_1$ spins, and bosonic spinons to describe the ${\bm S}_2$ spins.}
\label{fig:ancilla1}
\end{figure}
In this paper, we will obtain our effective theory of paramagnons and spinons by employing the recently developed ancilla qubit method to describe the FL* metal and the phases in its vicinity \cite{Zhang2020,Zhang2021,nikolaenko2021,Mascot22}. (We note the approach of Ref.~\cite{SS18a} in which the electron is not fractionalized, and the pseudogap metal is obtained by interactions between electrons and bosonic spinons---we will connect here to this approach in Section~\ref{sec:FL}. We also note other approaches \cite{Vlad17,Vlad21,Fabrizio19,Moreno22} using ancilla degrees of freedom.)
An earlier paper \cite{Mascot22} has shown that the ancilla method provides a good fit to the photoemission spectrum in the hole-doped cuprates in both the nodal and anti-nodal regions of the Brillouin, including a description of the momentum and energy dependence of the lineshapes in the anti-nodal region.

The ancilla method can be viewed as a simple and foolproof way of obtaining an effective low energy theory consistent with all symmetries, anomalies, and Luttinger relations.
The basic idea of this method is recalled in Fig.~\ref{fig:ancilla1}.
First, as shown in Appendix A of Ref.~\cite{nikolaenko2021}, we use an inverse Schrieffer-Wolff transformation, to transform the single-band Hubbard model to a model of non-interacting electrons coupled via Kondo coupling $J_K$ to a bilayer antiferromagnet of ancilla spins with rung-exchange $J_\perp$: at large $J_\perp$, the ancilla spins form rung singlets, and accounting for the virtual rung triplet excitations leads to a Hubbard model for the electrons with $U \sim J_K^2/J_\perp$.
The first ancilla layer has an antiferromagnetic Kondo coupling $J_K$ to the non-interacting electrons, while the second ancilla layer has an effective ferromagnetic Kondo coupling. The FL* phase is obtained when we assume that the antiferromagnetic Kondo coupling scales to strong coupling (as it does in the Kondo impurity problem) and dominates over $J_\perp$. Then we obtain the heavy Fermi liquid state of the Kondo lattice formed by the $c^\alpha$ layer and the ${\bm S}_1$ spins: this state has total hole density $1+p+1 = p \, (\mbox{mod} \, 2) $, and so yields the hole pockets of size $p$.
Meanwhile, the ${\bm S}_2$ spins with ferromagnetic Kondo couplings cannot be ignored: the ferromagnetic Kondo coupling scales to weak coupling, and so we can safely assume that the ${\bm S}_2$ spins decouple from the conduction electrons and form a spin liquid, which provides the neutral spinon excitations and the associated emergent gauge fields required in the FL* state. In summary, this theory of the pseudogap metal can be described by the slogan:
\begin{center}
\begin{tcolorbox}
[width=4in,  colback=white!95!black]
\begin{center}
Pseudogap Metal $=$\\
Kondo Lattice Heavy Fermi Liquid 
$\oplus$
Spin Liquid.
\end{center}
\end{tcolorbox}
\end{center}

\begin{figure}
\begin{center}
\includegraphics[width=5in]{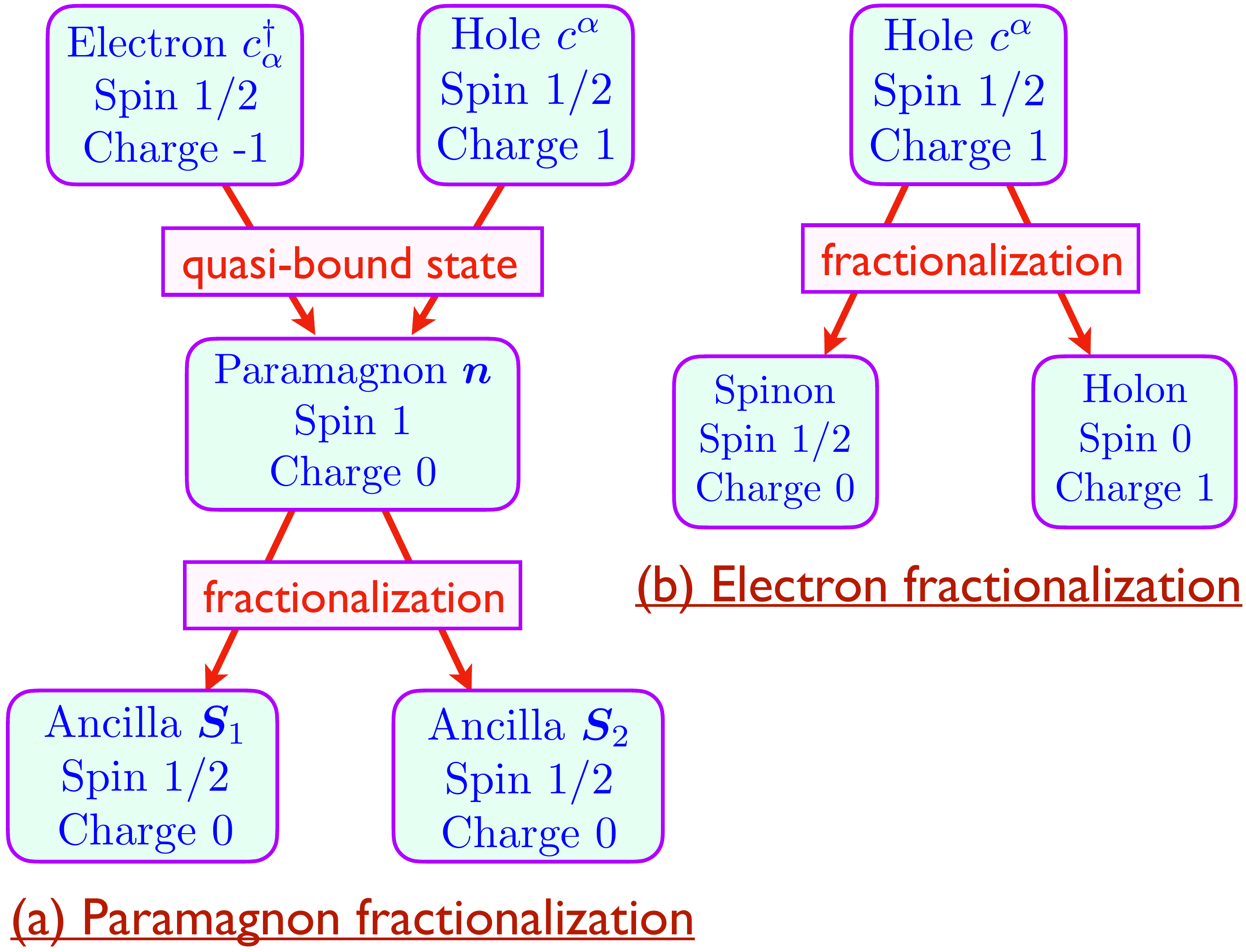}
\end{center}
\caption{Comparison of the paramagnon and electron fractionalization approaches. For the $t \gg J$ regime of the $t$-$J$ model description of the single-band Hubbard model the holon is unlikely to survive as a fractionalized excitation, and there is no clear experimental evidence for it. All charge carriers have spin $S=1/2$ in the paramagnon fractionalization approach employed in the present paper. We use the fermions $f^p$ to describe the ancilla spins ${\bm S}_1$, and bosons $Z^p$ to describe the ancilla spins ${\bm S}_2$.}
\label{fig:parafrac}
\end{figure}

A cartoon view of the parmagnon fractionalization approach is presented in Fig.~\ref{fig:parafrac}a.
The previous studies in the ancilla method \cite{Zhang2020,Zhang2021,nikolaenko2021,Mascot22} have used fermionic spinons to describe the spin liquid in the ${\bm S}_2$ spins. In the present paper, we shall use bosonic spinons to describe the spin liquid as that realized by the $\mathbb{CP}^1$ U(1) gauge theory \cite{AASB,rs1,rs2}, as in (\ref{i1}). Given the many dualities between the bosonic and fermionic spinon approaches \cite{DQCP3,SenthilSon,qpmbook}, we expect there is a non-perturbative mapping between the results obtained by the two approaches. But at the level of mean-field theory, and perturbative fluctuations, the results can be quite different, and much insight is gained by comparisons between them. Our study of the fermionic spinon dual description of the $\mathbb{CP}^1$ U(1) spin liquid is presented in a companion paper \cite{Christos:2023oru}: the dual has fermionic spinons moving in $\pi$ flux coupled to a SU(2) gauge field \cite{DQCP3}.

The mechanics by which the ancilla method delivers hole pocket Fermi surfaces of size $p$ has similarities to the 
``hidden'' fermion approaches with zeros in the electron Green's function \cite{ET02,ID03,Kotliar06,Georges06,Imada09,Imada10,Imada13,Imada16,Imada18,Imada19,Imada21,Imada21b,Fabrizio1,Fabrizio2,Fabrizio3} and ``YRZ''  \cite{YRZ,YRZ_rev} approaches. 
In all cases, the electron has a self-energy which is similar to the propagator of fermions in an auxiliary band; in the ancilla method, the auxilliary fermions reside on the first ancilla layer.
The spinon excitations arising from the second ancilla layer are not explicitly present in these earlier theories, but it has been argued \cite{SS17,Fabrizio1,Fabrizio2,Fabrizio3} that a similar role is played by the zeros of the electron Green's function. The zeros contribute a linear in $T$ specific heat and constant spin susceptibility \cite{Fabrizio1,Fabrizio2,Fabrizio3} in a manner similar to the mean-field theory of a spinon Fermi surface in a U(1)-FL* state. However, when we include the gauge fluctuations in the U(1)-FL* theory, the resulting $T^{2/3}$ specific heat is not captured by the theory of Green's function zeros.

Along with providing a description of the pseudogap metal as an FL* phase, the bosonic spinon approach of the present paper allows us to address confinement transitions of the fractionalized metal. 
\begin{figure}
\begin{center}
\includegraphics[width=5in]{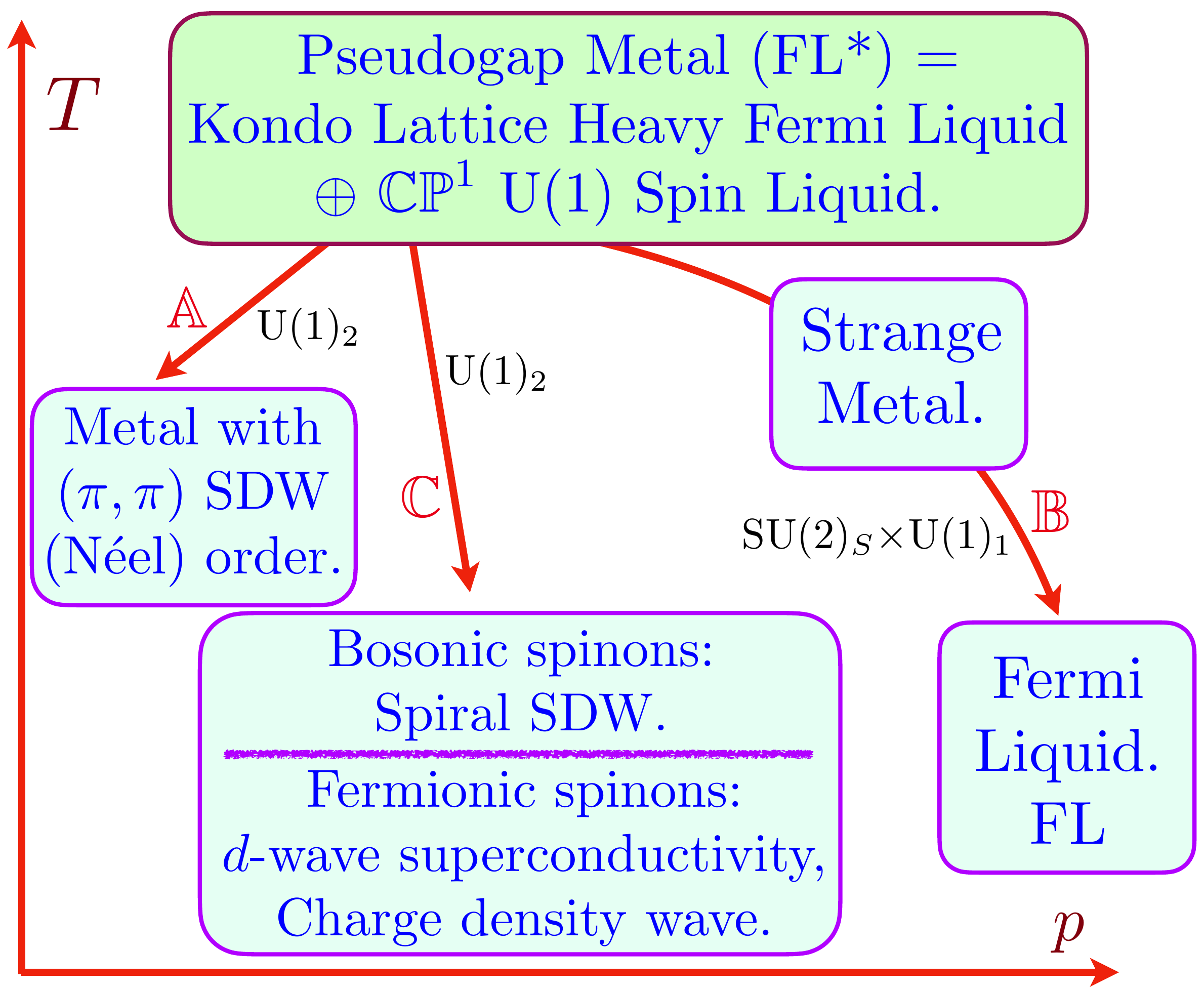}
\end{center}
\caption{Schematic of quantum phases in a temperature ($T$) and doping ($p$) phase diagram for the hole-doped cuprates. In this paper, we view the FL* pseudogap metal as the `parent', from which various low $T$ phases without fractionalization follow via confining or Higgs transition. The confinement/Higgs transitions along $\mathbb{A}$, $\mathbb{B}$, $\mathbb{C}$ are discussed in the text; also indicated are the primary gauge groups from Table~\ref{tab1} which drive these transitions. Along arrow $\mathbb{C}$, the bosonic spinon approach \cite{AASB,rs1,rs2} yields only a spiral SDW, as we shall see in Section~\ref{sec:stripes}. The dual fermionic spinon approach to the {\it same\/} spin liquid employs fermions moving in $\pi$ flux coupled to a SU(2) gauge field \cite{DQCP3}, and its confinement along $\mathbb{C}$ is described in Ref.~\cite{Christos:2023oru}, yielding $d$-wave superconductivity and charge density wave order.}
\label{fig:parent}
\end{figure}
It is relatively easy to reach the 
spin density wave (SDW) metallic state with N\'eel antiferromagnetic order at smaller $p$; we simply condense the $w^\alpha$ spinon \cite{CSS94}, as along arrow $\mathbb{A}$ in Fig.~\ref{fig:parent}. This simultaneously breaks spin and translational symmetries in the appropriate manner, and also Higgses out the U(1) photon so that no fractionalized excitations remain. If the $w^\alpha$ condensate is at zero wavevector, the spin density wave phase has conventional N\'eel order at wavevector $(\pi, \pi)$. However, the coupling to the charge carriers can also induce a $w^\alpha$ condensate at a non-zero wavevector (as along arrow $\mathbb{C}$ in Fig.~\ref{fig:parent}), leading to incommensurate spin density wave states, as we shall discuss in Section~\ref{sec:stripes}. We will describe the evolution of the Fermi surfaces from the FL* state to these SDW states. A significant result of our analysis is that the non-zero wavevector condensation of the $w^\alpha$ spinons of the $\mathbb{CP}^1$ U(1) spin liquid leads to a spiral SDW, and not to the collinear SDW associated with `stripe' states.

The transition from the FL* metal to the Luttinger-volume Fermi liquid phase (hereafter denoted FL) at larger $p$ cannot be described in such a facile manner. One of the purposes of this paper is to complete the phase diagram of this bosonic spinon approach to the FL* state, and also connect the FL phase to the FL* and SDW phases. 
The confinement transition from FL* to the Luttinger-theorem-obeying Fermi liquid (along arrow $\mathbb{B}$ in Fig.~\ref{fig:parent}) is quite involved, and proceeds via a rather exotic intermediate metallic phase D, as shown in Fig.~\ref{fig:pd}. 

We note in passing that there are numerous other approaches to the pseudogap metal ({\it e.g.\/} Refs.~\cite{Lee89,WenLee96,MeiWen12,Kaul07,ACL08,SS17,SS18b,Metzner2022}) which begin by fractionalizing the electron into a charge $+e$ spin $S=0$ `holon', and a spinon, as in Fig.~\ref{fig:parafrac}b. The main difficulty with these approaches is that they do not naturally lead to an FL* metal. In the case where the holon is a fermion, the mean-field description in such approaches leads to a `holon metal', with a Fermi surface of charge $+e$ spin $S=0$ quasiparticles. A Fermi surface of `holes' rather than `holons' can then be obtained by arguing that the holons form bound states with the spinons to yield charge $+e$ spin $S=1/2$ quasiparticles, as in the FL* metal. In practice, however, this binding process is difficult to carry out with any degree of control. Furthermore, there is no clear experimental evidence for the existence of spinless charge carriers in any energy regime in the cuprates. So we avoid such electron fractionalization approaches in the present paper, and will only fractionalize the paramagnon, as in (\ref{i1}). 

Section~\ref{sec:emerge} describes the structure of the gauge theory of the ancilla approach, as summarized in Table~\ref{tab1}. Section~\ref{sec:action} presents the effective action, obtained by imposing the symmetries in Table~\ref{tab1}. The mean-field phase diagram is obtained in Section~\ref{sec:pd} for the case of $(\pi, \pi)$ SDW order, along with results on the evolution of the Fermi surfaces across the quantum phase transitions. Section~\ref{sec:stripes} extends our results to SDW ordering at other wavevectors. 

\section{Emergent gauge structure}
\label{sec:emerge}

The ancilla approach has an intricate structure of gauge charge assignments, and the resulting gauge theory is the main tool used to derive the effective actions we shall work with below. This structure can be understood simply from rather general arguments, as we now show.

First, we have the U(1) gauge charges carried by the bosonic spinons $w^\alpha$ of the second ancilla layer in (\ref{i1}). We refer to this as U(1)$_2$.
See Table~\ref{tab2} below.

We will continue to use a fermionic spinon ($\psi^\alpha$) representation of the spins in the first ancilla layer
\beq
{\bm S}_1 = \frac{1}{2} \psi_\alpha^\dagger \, {\bm \sigma}^\alpha_{~\beta} \, \psi^\beta \label{i2}
\eeq
This introduces a U(1) gauge invariance, which we will denote U(1)$_1$. (Actually, the full gauge invariance of (\ref{i2}) is SU(2) \cite{Affleck88}, but we will always work with spin liquids in which the SU(2) is Higgsed down to U(1), and so we ignore this feature.) See Table~\ref{tab2} below.

Finally, we need an SU(2)$_S$ gauge field to impose the rung spin-singlet structure of the ancilla layers, induced by the large $J_\perp$. This is realized by transforming to a rotating reference frame in spin space \cite{sdw09}. The spin-singlet projection requires that we perform the same SU(2) rotation $\mathcal{R}$ in both ancilla layers. So we introduce fermions $f^p$ and bosons $Z^p$ ($ p = \pm$) by the transformation
\beq
\left( \begin{array}{c} \psi^\uparrow \\ \psi^\downarrow \end{array} \right) = \mathcal{R} \left( \begin{array}{c} f^+ \\ f^- \end{array} \right)
\quad ; \quad
\left( \begin{array}{c} w^\uparrow \\ w^\downarrow \end{array} \right) = \mathcal{R} \left( \begin{array}{c} Z^+ \\ Z^- \end{array} \right)\,. \label{i3}
\eeq
The fields $f^p$ and $Z^p$ now both carry a fundamental SU(2)$_S$ gauge charge. In addition, as is clear from (\ref{i3}), $f^p$ carries a U(1)$_1$ gauge charge, and $Z^p$ carries a U(1)$_2$ gauge charge. These charge assignments are summarized in Table~\ref{tab1}. The monopoles in U(1)$_2$ will carry Berry phases \cite{rs1,rs2}.
\begin{table}
    \centering
    \begin{tabular}{|c|c|c|c|c|c|c|}
\hline
Field & Statistics & U(1)$_1$ & U(1)$_2$ & SU(2)$_S$ & U(1)$_g$ & SU(2)$_g$  \\
\hline 
\hline
$c^\alpha$ & fermion & 0 & 0 & ${\bm 1}$ & 1 & ${\bm 2}$ \\
$f^p$ & fermion & 1 & 0 & ${\bm 2}$ & 0 & ${\bm 1}$ \\
$Z^p$ & boson & 0 & 1 & ${\bm 2}$ & 0 & ${\bm 1}$ \\
$\Phi_\alpha^p$ & boson & 1 & 0 & ${\bm 2}$ & -1 & $\bar{\bm 2}$ \\
$n^{a}$ & boson & 0 & 0 &${\bm 1}$ & 0 & ${\bm 3}$ \\
\hline
\end{tabular}
    \caption{Transformations of the main fields under gauge (U(1)$_1$, U(1)$_2$, SU(2)$_S$) and global (U(1)$_g$, SU(2)$_g$) symmetries. Listed are the charges under the U(1) symmetries, and the dimensions of the respresentations under the SU(2) symmetries. The effective action for the $c^\alpha$ (the gauge-invariant electrons), $f^p$ (spinons in the first ancilla layer), $Z^p$ (spinons in the second ancilla layer), and $\Phi_\alpha^p$ (Higgs field hybridizing the electrons with the first ancilla layer) is obtained by obeying these symmetries, and forms the basis of our results. The paramagnon field $n^a$ is a composite of the $Z^p$ and $\Phi_{\alpha}^p$ as defined in (\ref{i7}).}
    \label{tab1}
\end{table}
\begin{table}
    \centering
    \begin{tabular}{|c|c|c|c|c|c|c|}
\hline
Field & Statistics & U(1)$_1$ & U(1)$_2$ & SU(2)$_S$ & U(1)$_g$ & SU(2)$_g$  \\
\hline 
\hline
$\psi^\alpha$ & fermion & 1 & 0 & ${\bm 1}$ & 0 & ${\bm 2}$ \\
$w^\alpha$ & boson & 0 & 1 &  ${\bm 1}$ & 0 & ${\bm 2}$ \\
$H^{a \ell}$ & boson & 0 & 0 &${\bm 3}$ & 0 & ${\bm 3}$ \\
\hline
\end{tabular}
    \caption{As in Table~\ref{tab1}, for auxiliary fields used at intermediate stages.}
    \label{tab2}
\end{table}

These gauge charges combine to yield a (U(1)$_1 \times$U(1)$_2 \times$SU(2)$_S)/\mathbb{Z}_2$ gauge theory \cite{Zhang2020,Zhang2021}, some of whose phases we will study below. Note that the subsripts of the symmetry groups are just identifying labels, and do {\it not\/} refer to a Chern-Simons level.

In addition to the gauge charge assignments, we should also keep track of the global symmetries of charge and spin conservation. These we label as U(1)$_g$ and SU(2)$_g$. We present the global charge assignments of all the fields introduced so far in Tables~\ref{tab1} and \ref{tab2}.

The theory presented here requires one more boson which connects the global and gauge symmetries. This boson is the analog of the hybridization boson of the Kondo lattice, whose condensation yields the Kondo effect and the heavy Fermi liquid state \cite{Coleman84,ReadNewns,AuerbachLevin,LeeMillis}. Here, the required boson is a complex 4-component Higgs field $\Phi_\alpha^p$ \cite{Zhang2020,Zhang2021}. This hybridizes the fermions in the first ancilla layer with the electrons in the top layer, and so we have the operator correspondence
\beq
\Phi_\alpha^p \sim c_\alpha^\dagger f^p \label{i4}
\eeq
The gauge and global transformation properties of $\Phi_\alpha^p$ can be easily deduced from (\ref{i4}), and are listed in Table~\ref{tab1}. As discussed in Ref.~\onlinecite{Zhang2020}, the boson $\Phi_\alpha^p$ can be explicitly obtained from a Hubbard-Stratonovich transformation of the Kondo interaction between the electrons and the first ancilla layer.

Analogous to the Higgs field $\Phi_\alpha^p$, which connects the fermions $c^\alpha$ and $f^p$, it might seem we need another Higgs field to connect the SU(2)$_g$ spin space of $w^\alpha$ (as defined by (\ref{i1})) to the SU(2)$_S$ pseudospin space of $Z^p$. By taking the U(1)$_2$ gauge-invariant combination of the $Z^p$, we can introduce such a Higgs field $H^{a\ell}$ ($\ell = 1,2,3$, $a = x,y,z$) with $3 \times 3$ real components:
\beq
n^a = w^*_\alpha \, {\sigma}^{a \alpha}_{~~\beta} \, w^\beta \sim H^{a \ell} Z^*_p \, {\sigma}^{\ell p}_{~~p'} Z^{p'}\,, \label{i5}
\eeq
where $\sigma^\ell$ are also the Pauli matrices. Again ,the gauge and global charges for $H^{a\ell}$ can be deduced from Table~\ref{tab2}.
However, these symmetry properties also show that we can identify the Higgs field $H^{a \ell}$ as the `square' of the Higgs field $\Phi_\alpha^p$
\beq
H^{a \ell} = \Phi_\alpha^p \, \Phi^{\ast \beta}_{p'} \, \sigma^{a\alpha}_{~~\beta} \sigma^{\ell p'}_{~~p}\,. \label{i6}
\eeq
Consequently, we will not need to include $H^{a \ell}$ as an independent field in our considerations, and just identify it as in (\ref{i6}).
Also, we can combine (\ref{i5}) and (\ref{i6}) to write
\beq
n^a \sim  \Phi_\alpha^p \, \Phi^{\ast \beta}_{p'} \, \sigma^{a\alpha}_{~~\beta} \left[ Z^\ast_p Z^{p'} - \tfrac{1}{2}\, \delta_{p}^{p'} \, Z_q^\ast Z^q\right]\,, \label{i7}
\eeq
which relates the paramagnon field $n^a$ to the Higgs field $\Phi_\alpha^p$ and the spinons $Z^p$.

The remainder of the paper will derive an effective action for the electrons $c^\alpha$, the fermionic spinons $f^p$, the bosonic spinons $Z^p$, and the Higgs field $\Phi_\alpha^p$. This action can largely be deduced from the gauge and symmetry properties listed in Table~\ref{tab1}. Our results for the phase diagram and the properties of the phases will follow from this effective action.

\section{Effective action}
\label{sec:action}

Our primary assumption is that the structure of the phase diagram is determined primarily by the dynamics of the Higgs fields $Z^p$ and $\Phi_\alpha^p$, and the associated U(1)$_1$, U(1)$_2$, and SU(2)$_S$ gauge fields. In all our discussion here, we will not write out the gauge fields explicitly, as they can be included from the requirements of gauge invariance in a familiar manner. The fermionic matter fields $c^\alpha$ and $f^p$ are also important, and their couplings to the Higgs fields are determined, as usual, by the restrictions of gauge invariance: these couplings then modify the Fermi surfaces, and determining the Fermi surface evolution will be an important focus of our study. 

We begin by writing down the form of the Higgs potential, whose minima will determine the structure of the mean-field phase diagram. From Table~\ref{tab1} we have
\bea
V(Z, \Phi) &=& s_1\, \Phi_{\alpha}^p \Phi^{\ast \alpha}_p + u_1 \left[ \Phi_{\alpha}^p \Phi^{\ast \alpha}_p  \right]^2 +v_1 \, \Phi_{\alpha}^p \Phi_\beta^q \Phi^{\ast \alpha}_q \Phi^{\ast \beta}_p \nonumber \\
&+& s_2 \, Z^\ast_p Z^p +u_2  \left[ Z^\ast_p Z^p \right]^2 \nonumber \\
&+& w_1 \, Z^\ast_p Z^p \Phi_{\alpha}^q \Phi^{\ast \alpha}_q + w_2 \, Z^\ast_p Z^q \Phi_{\alpha}^p \Phi^{\ast \alpha}_q + \ldots \label{a1}
\eea
A variety of minima are possible from such a potential as the `masses' $s_{1,2}$ are varied, but we will limit ourselves to the regime where 
the minima can be related by gauge and global rotations to
\beq
\langle \Phi_{\alpha}^p \rangle = \bar\Phi \, \delta_\alpha^p \quad, \quad \langle Z^p \rangle = (\bar Z \, \delta^{p}_+ + \bar Z^\ast \,\delta^{p}_-)/\sqrt{2}\,. \label{a2}
\eeq
With $\bar{Z} = 0$ and $\bar{\Phi} \neq 0$, such a minimum yields the FL* state which breaks no global symmetries, and which has been studied in some detail in previous work \cite{Zhang2020,Zhang2021,nikolaenko2021,Mascot22}. Our purpose here is to study the remainder of the phase diagram when $\bar{Z}$ is also allowed to be non-zero.

There are also spatial and temporal gradient terms in $Z^p$ and $\Phi_{\alpha}^p$. But we refrain from writing them out explicitly because they have a familiar form dictated by gauge invariance, and are not used in the analysis of the present paper. Similarly, there are Maxwell terms for the gauge fields, and monopole Berry phases for the U(1)$_2$ gauge field in the second ancilla layer, which we do not present here \cite{rs1,rs2,senthil1,senthil2,DQCP3,DQCP4}.

Finally, let us discuss the fermionic sector, which can also have a significant influence on the fate of fluctuations.
As in previous work \cite{Zhang2020,Zhang2021,nikolaenko2021,Mascot22}, we have the dispersions of the electrons $c^\alpha$ in the physical layer, and the fermions $f^p$ in the first ancilla layer, along with their hybridization (Yukawa) coupling to $\Phi_\alpha^p$:
\begin{align}
H_{cf}^a = -  \sum_{i,j} t_{ij} c^\dagger_{i\alpha} c_{j}^{\alpha}+  \sum_{i,j} t_{1,ij} f^\dagger_{ip} f_{j}^{p}+ \sum_{i} \left( 
\Phi_{\alpha}^p f^\dagger_{ip} c_{i}^{\alpha}+ \Phi_p^{*\alpha} c^\dagger_{i\alpha} f_{i}^p \right) \,. \label{a3}
\end{align}
Using the $\Phi$ condensate in (\ref{a2}), $H_{cf}^a$ then yields the fermion dispersion in the FL* phase. Such dispersions were compared with photoemission observations in Ref.~\onlinecite{Mascot22}, and were able to describe observations well in both the nodal and anti-nodal regions of the Brillouin zone, 
and at and away from the Fermi surface. This comparison with data also allowed the determination of the hopping parameters and chemical potentials in 
$t_{ij}$ and $t_{1,ij}$, and the hybridization $\bar\Phi$.

The symmetries in Table~\ref{tab1} allow a number of additional couplings between the Higgs fields and the fermions (which were not considered in Ref.~\onlinecite{Mascot22})
\begin{align}
H_{cf}^b = & - J_s \sum_i \eta_i \, n^a_i \, c_{i \alpha}^\dagger \sigma^{a\alpha}_{~~\beta} c_{i}^\beta + J_\perp \sum_i \eta_i  \, f^{\dagger}_{ip} \sigma^{\ell p}_{~~p'} f_{i}^{p'}  \, Z^*_{iq}\sigma^{\ell q}_{~~q'} Z_i^{q'}+
\nonumber \\
& +J_3 \sum_i\left( \eta_i \, f^{\dagger}_{ip} \sigma^{\ell p}_{~~\alpha} \Phi^{ \alpha}_{\beta} c_{i}^{\beta}  \, Z^*_{iq}\sigma^{\ell q}_{~~q'} Z_i^{q'}+h.c.\right) + \lambda \sum_i \Phi_{\alpha}^p \sigma^{a \alpha}_{~~\beta} \Phi^{\ast \beta}_{p} \, c_{i \gamma}^\dagger \sigma^{a\gamma}_{~~\delta} c_{i}^\delta
\,, \label{a4}
\end{align}
where 
\beq
\eta_i \equiv (-1)^{i_x + i_y} \label{a5}
\eeq
is the staggering factor needed because $Z^p$ describes N\'eel order in the second ancilla layer. More formally, the $w^\alpha$ and $Z^p$ transform non-trivially under lattice symmetries \cite{Balents:2006wm}, and this implies the presence of $\eta_i$.
This N\'eel order is coupled to the electrons $c^\alpha$ via $J_s$ (where $n^a$ is related to $Z^p$ and $\Phi_\alpha^p$ as in (\ref{i7})), and to the first layer of ancilla fermions via $J_\perp$. The $\lambda$ term is proportional to the ferromagnetic moment, and so vanishes in mean-field theory in all the phases considered here.

\section{Phase diagram}
\label{sec:pd}

We follow these steps to describe the phase diagram:
\begin{itemize}
\item Determine the mean-field phase diagram by minimizing the Higgs potential in (\ref{a1}) to obtain the values of $\bar\Phi$ and $\bar Z$.
\item Insert the values of $\bar\Phi$ and $\bar Z$ into $H_{cf}^a + H_{cf}^b$ in (\ref{a3}) and (\ref{a4}), and then compute the mean-field dispersions of the fermions and their Fermi surfaces.
\item Analyze gauge fluctuations in all the phases and phase transitions so obtained.
\end{itemize}

\begin{figure}
\begin{center}
\includegraphics[width=4.5in]{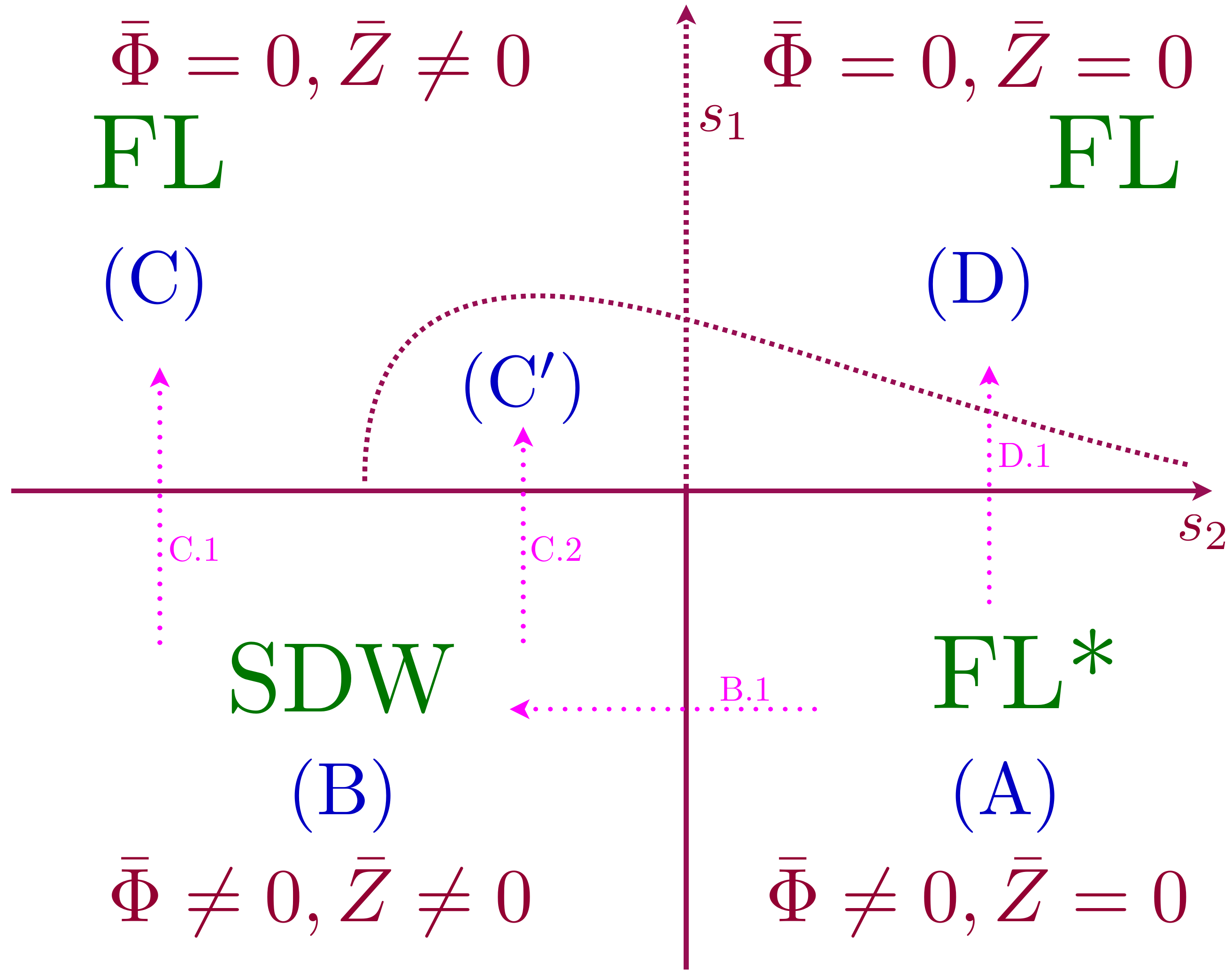}
\end{center}
\caption{Mean-field phase diagram of (\ref{p1}) for $w_1 + w_2 = 0$. The fermion spectrum and gauge fluctuations in the phases are described in the A,B,C,D subsections of Section~\ref{sec:pd}. The Fermi surface evolution along the dotted arrows is described in the labeled subsections of Section~\ref{sec:pd}. Region C$^\prime$ has a large $c^\alpha$ Fermi surface (as in the FL phase), along with small pocket ghost Fermi surfaces and  emergent (U(1)$_1 \times$U(1)$_d)/\mathbb{Z}_2$ gauge fields. The boundary of C$^\prime$ is vertical and $s_1$-independent in mean-field theory, but we have sketched a curved boundary to connect with the possible ghost Fermi surfaces in region D. Most of region D is also expected to be FL, apart from at or near the transition to FL*, where a large ghost Fermi surface may appear along with emergent (U(1)$_1 \times$SU(2)$_S)/\mathbb{Z}_2$ gauge fields \cite{Zhang2020}. The quantum phase transition(s) along arrow D.1 has been discussed in earlier work \cite{Zhang2020,Zhang2021} }
\label{fig:pd}
\end{figure}

We begin by describing the first step above, the mean-field theory of (\ref{a1}). With the ansatz (\ref{a2}), this becomes the standard Landau theory of tetracritical and bicritical points \cite{Nelson74,Nelson76}, with the Landau potential
\begin{equation}
     V(Z,\Phi)=s_1 |\bar\Phi|^2+(u_1+v_1) |\bar\Phi|^4+s_2 |\bar Z|^2+u_2 |\bar Z|^4+(w_1+w_2)|\bar Z|^2 |\bar\Phi|^2 \label{p1}
\end{equation}
We will not work out the phase diagram of (\ref{p1}) here, as it is identical to that in early works  \cite{Nelson74,Nelson76}. In the interests of simplicity, we focus on the case $w_1 + w_2=0$, when the phase diagram takes the very simple form in Fig.~\ref{fig:pd}.
 The mean-field theory yields 4 phases, A, B, C, D, separated by phase boundaries at $s_1=0$ or $s_2=0$. These phases are discussed below in the correspondingly named subsections. Upon including gauge fluctuations, phases C and D ultimately become conventional Fermi liquids (FL) with a single large Fermi surface of the $c^\alpha$ fermions; so phases C and D can be smoothly connected without an intervening quantum phase transition, and this is indicated by making the line $s_2=0$, $s_1>0$ a dashed line. However, within regions C and D, below the curved dotted line, there is the possibility of additional `ghost' Fermi surfaces of the $f^p$ fermions \cite{Zhang2020,Zhang2021}; these ghost Fermi surfaces will be small for $s_2<0$, and large for $s_2>0$.

\subsection{FL*}

This phase has $\bar\Phi \neq 0$, $\bar Z=0$. The $\Phi_\alpha^p$ condensate fully Higgses the SU(2)$_S$ and U(1)$_1$ gauge fields, but U(1)$_2$ gauge field remains potentially deconfined. 

With $\bar Z=0$, the situation here is as described in earlier papers \cite{Zhang2020,Zhang2021,nikolaenko2021,Mascot22}, and also along the eventual transition to a Fermi liquid in region D. So we obtain small hole pocket Fermi surfaces of size $p$, a deconfined U(1)$_2$ gauge field. 

The bosonic spinon description of the spin liquid in the second ancilla layer can make a potential difference here from the earlier work: the U(1)$_2$ spin liquid can have a monopole-induced confinement to a valence bond solid at some large length scale. Alternatively, a stable $\mathbb{Z}_2$ spin liquid can appear here \cite{senthil1,senthil2,DQCP3,DQCP4}.

\subsection{SDW}

This phase has $\bar\Phi \neq 0$, $\bar Z \neq 0$. Now the $\Phi_\alpha^p$ and $Z^p$ condensates fully Higgs all the SU(2)$_S$, U(1)$_1$, and U(1)$_2$ gauge fields, and there are no deconfined gauge charges. However, the presence of both Higgs condensates implies that $\langle {\bm n} \rangle \neq 0$ from (\ref{i7}), and the global symmetry SU(2)$_g$ is broken, implying the presence of SDW order. 

\subsubsection{Fermi surfaces from FL* to SDW}
\label{flstar-sdw}

Here we follow arrow $\mathbb{A}$ in Fig.~\ref{fig:parent}, or  arrow B.1 in Fig.~\ref{fig:pd}. To compute Fermi surfaces we start from mean-field Hamiltonian in the reduced Brillouin zone $H=\sum_{\kk} \psi^{\dagger}_{\kk} H^{\phantom{\dagger}}_{\kk} \psi^{\phantom{\dagger}}_{\kk}$, where $\psi_{\kk}=(c_{\kk}, c_{\kk+\qp},f_{\kk},f_{\kk+\qp})$, $\qp=(\pi,\pi)$ and (with $\bar\Phi$, $\bar Z$ real)
\begin{equation}
 H_{\kk}=
\left(
\begin{array}{cccc}
 \ec{\kk}  & J_s \bar Z^2\bar\Phi^2 & \bar\Phi &0  \\
J_s \bar Z^2 \bar\Phi^2 & \ec{\kk+\qp}& 0 &\bar\Phi\\
 \bar\Phi & 0 & \ef{\kk} &  J_\perp \bar Z^2 \\
 0 & \bar\Phi & J_\perp \bar Z^2&\ef{\kk+\qp} \\
\end{array}
\right)\,. \label{eq:mat4}
\end{equation}
With the input of the variations in the values of $\bar \Phi$ and $\bar Z$ across the phase diagram of Fig.~\ref{fig:pd}, this Hamiltonian describes the mean-field evolution of the Fermi surfaces across all the phases.
The eigenvalues of $H_{\kk}$ cannot be found analytically for nonzero $\bar Z$ and $\bar\Phi$ but it is easy to diagonalize the Hamiltonian numerically. We choose $J_s = J_\perp=1$, and tight-binding parameters consistent with experimental ARPES observations in Bi2201 \cite{He2011}: 
\begin{align}
    \ec{\kk}=&-2t(\cos k_x+\cos k_y)-4 t' \cos k_x \cos k_y-2t''(\cos 2 k_x+\cos 2 k_y)\nonumber\\
    &-4t'''(\cos 2 k_x \cos k_y+\cos 2 k_y \cos k_x)-\mu_c \, , \label{ecdef}
\end{align}
 where $t=0.22, t'=-0.034, t''=0.036, t'''=-0.007, \mu_c=-0.24$ and
 \begin{align}
     \ef{\kk}=2t_1(\cos k_x+\cos k_y)+4 t_1' \cos k_x \cos k_y+2t_1''(\cos 2 k_x+\cos 2 k_y)-\mu_f \, ,\label{efdef}
 \end{align}
where $t_1=0.1, t_1'=-0.03, t_1''=-0.01, \mu_f=0.009$. Chemical potentials will vary in order to satisfy constraints $\langle c^{\dag}_{i\alpha}c_i^\alpha\rangle=(1-p)/2$ and $\langle f^{\dag}_{i p}f_i^p\rangle=1/2$, while we will use the same tight-binding parameters in the rest of the paper. We also compute spectral weights, by taking imaginary part of retarded Green's function with finite imaginary broadening $\delta=0.01$. The spectral weight, contrary to a Fermi surface, is directly measured in ARPES experiments.

     \begin{figure}[h]
\begin{minipage}[h]{0.4\linewidth}
  \center{\includegraphics[width=1\linewidth]{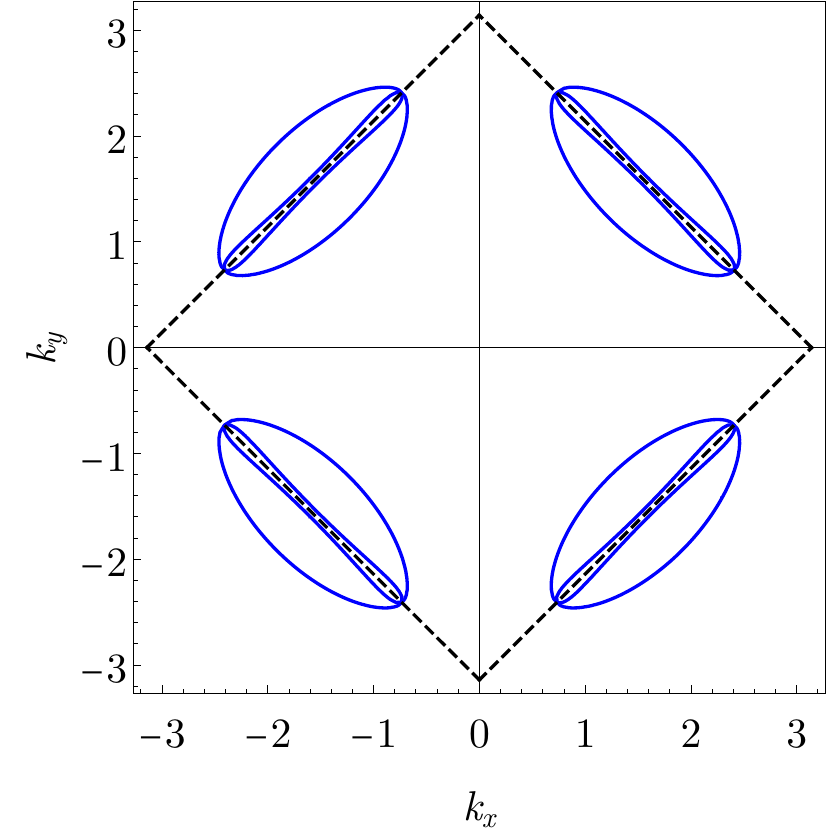}}
  
  \end{minipage} 
  \begin{minipage}[h]{0.4\linewidth}
  \center{\includegraphics[width=1.45\linewidth]{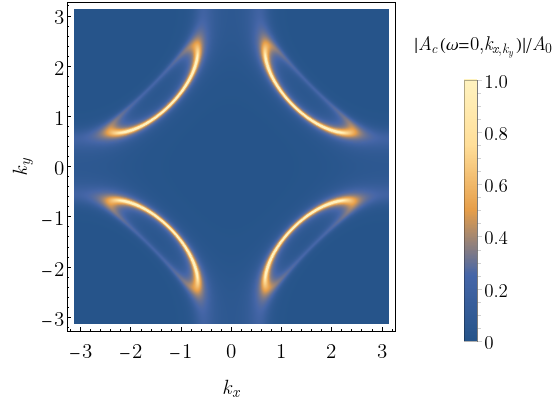}}
 
  \end{minipage} 
\caption{Fermi surface and spectral weight in the FL$^*$ phase. Parameters: $\bar\Phi=0.09, \bar Z=0.0$, $\mu_c=-0.243, \mu_f=0.009$. }
\label{fig:fermi_surface_1}
\end{figure}
Fig.~\ref{fig:fermi_surface_1} shows the Fermi surface and spectral weight in the FL$^*$ phase. There are eight hole pockets instead of four since the Brillouin zone is shrunk by two in our basis, but the spectral weight only sees four hole pockets. 

\begin{figure}[h]
\begin{minipage}[h]{0.4\linewidth}
  \center{\includegraphics[width=1\linewidth]{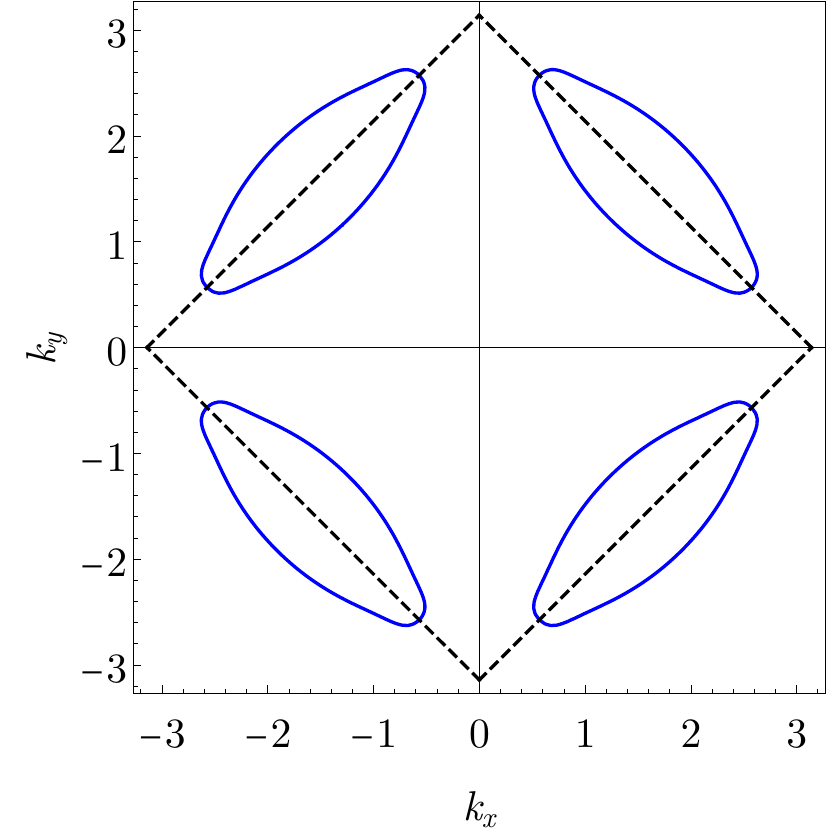}}
  
  \end{minipage} 
  \begin{minipage}[h]{0.4\linewidth}
  \center{\includegraphics[width=1.45\linewidth]{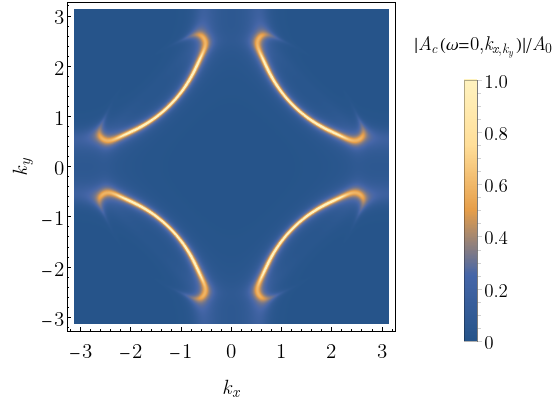}}
 
  \end{minipage} 
\caption{Fermi surface and spectral weight in the SDW phase. Parameters: $\bar\Phi=0.09, Z=0.2$, $\mu_c=-0.237, \mu_f=0.006$. }
\label{fig:fermi_surface_2}
\end{figure}

As we move into the SDW phase along B.1 line, eight hole pockets turn into four hole pockets, as shown in Fig.~\ref{fig:fermi_surface_2}. Unlike the FL* pockets in Fig.~\ref{fig:fermi_surface_1}, these pockets are symmetric with respect to the boundaries of the reduced Brillouin zone (black dashed line). 
Moreover, their area enclosed in the original Brillouin zone has doubled \cite{Kaul07}. These are features which should be possible to detect in experiment.

The transition is described by a $\mathbb{CP}^1$ field theory for the U(1)$_2$ gauge field coupled to $Z^p$, along with a spectator bands of fermions neutral under 
U(1)$_2$; recall that the monopoles in U(1)$_2$ do carry Berry phases, and this allows deconfined criticality \cite{senthil1,senthil2}.

\subsection{FL}
\label{sec:FL}

This phase has $\bar\Phi = 0$, $\bar Z \neq 0$. The $Z^p$ condensate breaks the (SU(2)$_S \times$U(1)$_2)/\mathbb{Z}_2$ gauge symmetry down to a diagonal U(1) symmetry which we refer to as U(1)$_d$: this is the linear combination of U(1)$_2$ and the $x$ component of SU(2)$_S$ which leaves the second equation in (\ref{a2}) invariant (for real  $\bar Z$).
The U(1)$_1$ gauge field is also potentially deconfined. 
The $f^p$ Fermi surface can be gapped via the $J_\perp$ coupling in (\ref{a4}) and (\ref{eq:mat4}) provided the $Z^p$ condensate is large enough. We assume the $f^p$ fermions are gapped for now, and consider the situation with gapless $f^p$ excitations below.
With the $f^p$ fermions gapped, the Polyakov mechanism of monopole proliferation can confine (U(1)$_1\times$U(1)$_d)/\mathbb{Z}_2$ gauge fields, but we do need to consider the monopole Berry phases to determine the scales over which deconfinement can survive \cite{rs1,rs2,senthil1,senthil2}. Upon confinement, the Higgs field $\Phi^p_\alpha$ is no longer an elementary excitation in the FL phase. The gauge-neutral 2-particle bound state of $\Phi^p_\alpha$ turns into the paramagnon via (\ref{i7}). For the explicit form of (\ref{i7}) here, it is useful to orient the $Z^p$ condensate in the $z$ direction by replacing (\ref{a2}) by $\langle Z^p \rangle = \bar Z \, \delta^{p}_+$; then (\ref{i7}) becomes
\bea
{\bm n} &\sim& \Phi_\alpha^+ {\bm \sigma}^{\alpha}_{~~\beta} \Phi^{\ast \beta}_{+} -  \Phi_\alpha^- {\bm \sigma}^{\alpha}_{~~\beta} \Phi^{\ast \beta}_{-} \nonumber \\
{\bm m} &\sim& \Phi_\alpha^+ {\bm \sigma}^{\alpha}_{~~\beta} \Phi^{\ast \beta}_{+} +  \Phi_\alpha^- {\bm \sigma}^{\alpha}_{~~\beta} \Phi^{\ast \beta}_{-}\,. \label{nm}
\eea
We have also noted the form of the ferromagnetic order parameter ${\bm m}$, which follows from the $\lambda$ coupling in (\ref{a4}).
The $\Phi^p_\alpha$ now carry charge $p$ under U(1)$_d$, and charge 1 under U(1)$_1$.
We can realize a theory without ferromagnetism, ${\bm m} \approx 0$, by condensing $\varepsilon^{\alpha\beta} \Phi_{\alpha}^+ \Phi_{\beta}^-$ so that 
$\Phi_\alpha^- \sim \varepsilon_{\alpha\beta} \Phi_{+}^{\beta \ast}$. This condensate higgses U(1)$_1$, but there remains the possibility that
 U(1)$_d$ is deconfined over a signficant length scale, in which case we should consider the theory with fractionalized 
$\Phi_\alpha^+$ excitations: such a theory reduces to that considered in Section III of Ref.~\onlinecite{SS18a} with the bosonic spinon $z_\alpha^\ast$ of that paper corresponding to our $\Phi_\alpha^+$. However note that the spinons of $\Phi_\alpha^+$ represent spin fluctuations on both ancilla layers, and so the Berry phases of the monopoles in U(1)$_d$ {\it cancel\/} between the contributions of the two layers, and do not suppress confinement \cite{senthil1,senthil2}.
Once U(1)$_d$ confines, we obtain the usual Hertz theory \cite{hertz} of a paramagnon ${\bm n} \sim \Phi_\alpha^+ {\bm \sigma}^{\alpha}_{~~\beta} \Phi^{\ast \beta}_{+}$ coupled to large Fermi surface. The deconfined theory in terms of the $\Phi_\alpha^+$ has no monopoles/hedgehogs and so only includes orientational fluctuations of the SDW order, whereas the ${\bm n}$ theory also allows amplitude fluctuations.

For a smaller $Z^p$ condensate, the $f^p$ fermions can be gapless because pocket $f^p$ ghost Fermi surfaces will survive, and we have indicated this region of the phase diagram as C$^\prime$ in Fig.~\ref{fig:pd}. With gapless $f^p$ fermions, the Polyakov mechanism for confinement is suppressed \cite{SSLee_spinon}.
The $f^p$ fermions have gauge charges $p=\pm$ under U(1)$_d$, and the same gauge charge under U(1)$_1$: consequently there is a near-cancellation of attractive and repulsive forces \cite{Zhang2021}, and it is possible that the pocket $f^p$ Fermi surfaces will avoid a pairing instability.
If the pairing instability does occur, the ancilla layers become trivial, and we obtain a conventional FL state---the (U(1)$_1\times$U(1)$_d)/\mathbb{Z}_2$ gauge symmetry is Higgsed down to U(1)$_d$, and the fermion gap will 
lead to the U(1)$_d$ confinement discussed above. 

\subsubsection{Fermi surfaces from SDW to FL}
\begin{figure}[h]
\begin{minipage}[h]{0.4\linewidth}
  \center{\includegraphics[width=1\linewidth]{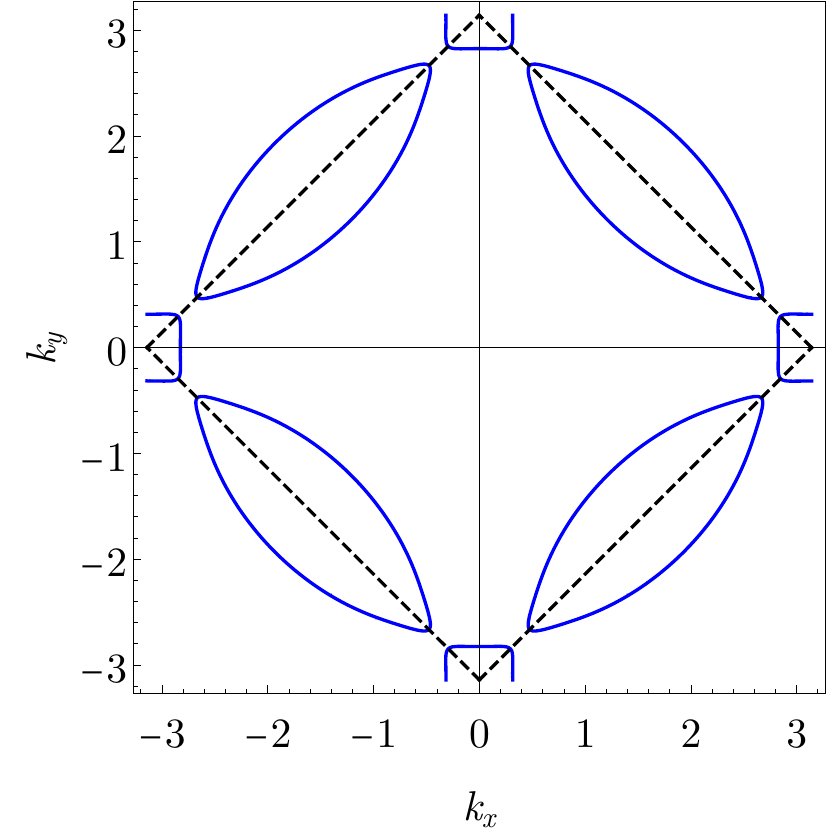}}
  \end{minipage} 
  \begin{minipage}[h]{0.4\linewidth}
  \center{\includegraphics[width=1.45\linewidth]{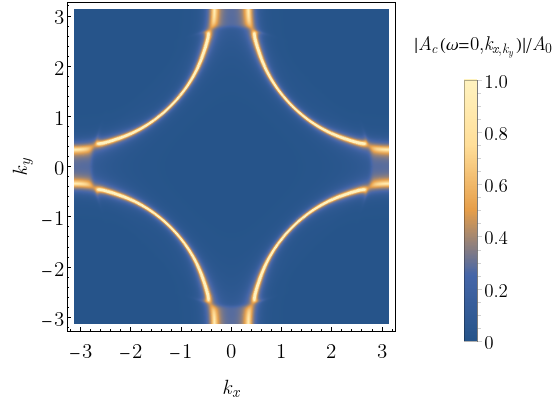}}
  \end{minipage} 
\caption{Fermi surface and spectral weight in the SDW phase, close to FL phase. Parameters: $\Phi=0.03, Z=0.2$, $\mu_c=-0.210, \mu_f=0.062$. }
\label{fig:fermi_surface_3}
\end{figure}

The transition from SDW to Fermi liquid (along C.1 line) is shown in Figure \ref{fig:fermi_surface_3}. As we move closer to FL phase electron pockets appear at the antinodal points and Fermi arcs evolve into a usual Fermi surface of a Fermi liquid.

The $f^p$ electron Fermi surface is gapped, and so as discussed above, the transition is a conventional transition \cite{hertz} from SDW to FL, with the $J_s$ term in (\ref{a4}) coupling to the SDW order parameter, the paramagnon ${\bm n}$ defined by (\ref{nm}).

\subsubsection{Fermi surfaces from SDW to C'}
\begin{figure}[h]
\begin{minipage}[h]{0.4\linewidth}
  \center{\includegraphics[width=1\linewidth]{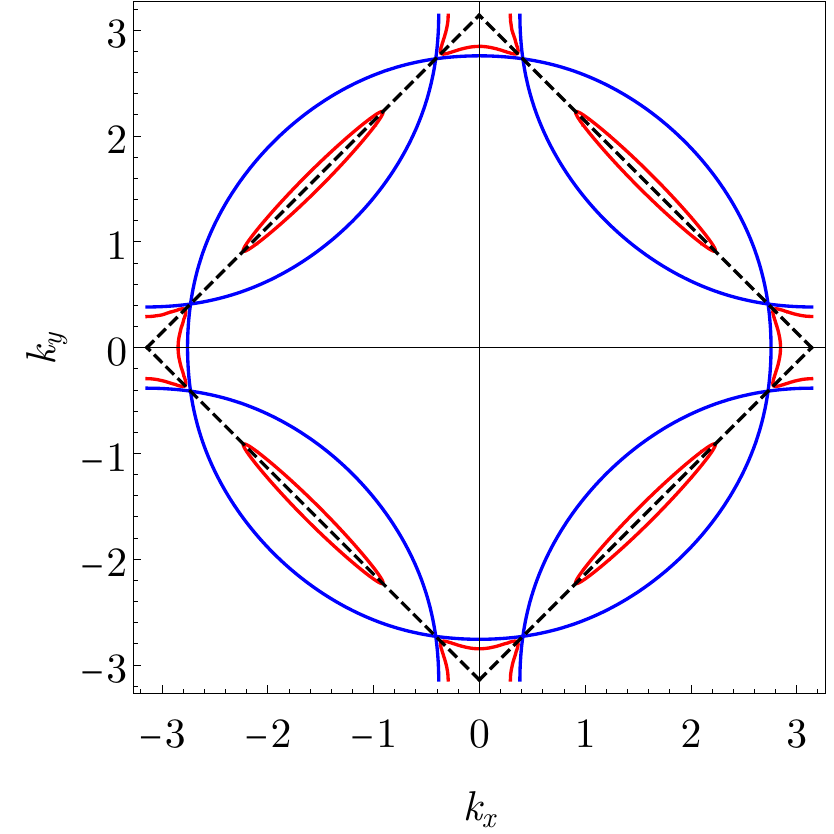}}
  \end{minipage} 
  \begin{minipage}[h]{0.4\linewidth}
  \center{\includegraphics[width=1.45\linewidth]{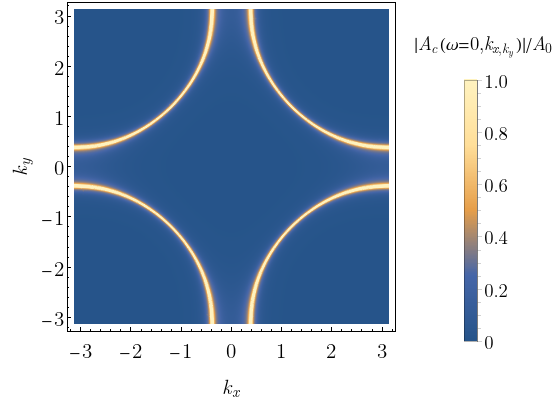}}
  \end{minipage} 
\caption{Fermi surface and spectral weight in the C' phase. The blue line corresponds to   $c^\alpha$ Fermi surface and red line corresponds to $f^p$ ghost Fermi surface. Parameters: $\Phi=0.0, Z=0.1$, $\mu_c=-0.213, \mu_f=0.065$. }
\label{fig:fermi_surface_4}
\end{figure}

The transition between SDW and C' phase happens when $\bar Z$ is smaller and the Fermi surface of $f^p$ fermions is not gapped. Figure \ref{fig:fermi_surface_4} shows $c^\alpha$ fermion Fermi surface (blue line), and $f^p$ fermion Fermi surface (red line). However, spectral weight lies only on the physical electron Fermi surface.

Assuming the (U(1)$_1\times$U(1)$_d)/\mathbb{Z}_2$ gauge fields are deconfined, the transition is described by a gauge theory for the $\Phi^p_\alpha$ and the fermions; for $\langle Z^p \rangle = \bar Z \, \delta^{p}_+$, the $\Phi^p_\alpha$ and $f^p$ carry gauge charges $p$ under U(1)$_d$ and $1$ under U(1)$_1$. The monopoles in both U(1)'s are suppressed by the $f^p$ Fermi surfaces, and they do not carry Berry phases because the $\Phi^p_\alpha$ represent spin fluctuations in a bilayer antiferromagnet.

\subsection{Ghost Fermi surfaces}

\subsubsection{Fermi surfaces from FL* to D}

\begin{figure}[h]
\begin{minipage}[h]{0.45\linewidth}
  \center{\includegraphics[width=1.2\linewidth]{spectral_weight_c_phi_0.09_z_0.00.png}}
  \end{minipage} 
  \begin{minipage}[h]{0.45\linewidth}
  \center{\includegraphics[width=1.2\linewidth]{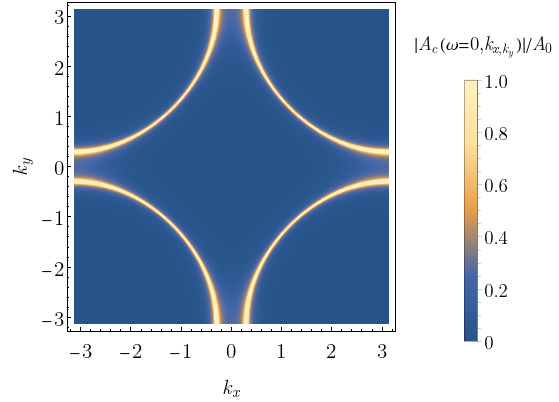}}
  \end{minipage} 
\caption{Spectral weight in the FL$^*$ and D phase. Parameters: $\Phi=0.09, Z=0.0$, $\mu_c=-0.24, \mu_f=0.009$. $\Phi=0$ in the D phase. }
\label{fig:fermi_surface_5}
\end{figure}

 The transition between FL$^*$ and D was studied in previous works \cite{Zhang2020,Zhang2021,nikolaenko2021,Mascot22} and describes an emergence of hole pockets from the full Fermi surface, see Figure \ref{fig:fermi_surface_5}.

\section{Incommensurate spin density waves}
\label{sec:stripes}

This section will explore the possibility that the SDW phase in Fig.~\ref{fig:pd} has incommensurate spin correlations, as indicated as a possibility along arrow $\mathbb{C}$ in Fig.~\ref{fig:parent}. This is motivated by the numerous observations of incommensurate spin order in the underdoped regime of the La-based cuprates. 

We approach the SDW phase B from the FL* phase A. Both these phases have the Higgs condensate $\left\langle \Phi^p_\alpha \right\rangle \neq 0$, while only the SDW phase has $\left\langle Z^p \right\rangle \neq 0$. Our operating assumption is that the coupling to the Fermi pockets of the FL* phases, leads the $Z^p$ spinons to condense at an incommensurate wavevector. In the remaining discussion in this section we will not distinguish between the SU(2)$_S$ indices and the SU(2)$_g$ indices because they are identified by the diagonal $\Phi^p_\alpha$ condensate in (\ref{a2}). 
Moreover, the spinons $w^\alpha$ are identified with $Z^p$ by (\ref{i5}) and (\ref{i6}).
So we will denote the spinons  by $Z^\alpha$ in this section, and (\ref{i7}) becomes
\beq
n^a =  Z_\alpha^\ast \, \sigma^{a\alpha}_{~~\beta} Z^\beta\,. \label{i7a}
\eeq

We are interested in a state in which the spinon condensate in (\ref{a2}) is replaced by
\beq
\left \langle Z^\alpha (\vec{r}) \right\rangle = \bar{Z}_1^{\alpha} e^{i \vec{q}_1 \cdot \vec{r}} + \bar{Z}_2^{\alpha} e^{i \vec{q}_2 \cdot \vec{r}} +  \bar{Z}_3^{\alpha} e^{-i \vec{q}_1 \cdot \vec{r}} + \bar{Z}_4^{\alpha} e^{-i \vec{q}_2 \cdot \vec{r}}\label{i8}
\eeq
where $\vec{q}_1$ and $\vec{q}_2$ incommensurate wavevectors related by square lattice symmetries. Thus, we could have $\vec{q}_1 = (\kappa, 0)$ and $\vec{q}_2 = (0, \kappa)$, or we could have $\vec{q}_1 = (\kappa, \kappa)$ and $\vec{q}_2 = (-\kappa, \kappa)$, where $\kappa$ is some small incommensurate wavevector. Inserting (\ref{i8}) into (\ref{i7a}), we see that the possible ordering wavevectors of the SDW order are
\beq
(\pi, \pi) \quad ; \quad (\pi, \pi) \pm 2\vec{q}_1 \quad ; (\pi, \pi) \pm 2 \vec{q}_2 \quad ; \quad (\pi, \pi) \pm \vec{q}_1 \pm \vec{q}_2\,. \label{i9}
\eeq
Of significant importance is the fact that there remains an SDW ordering at the commensurate wavevector $(\pi, \pi)$ with weight proportional to
\beq
\bar{Z}_{1\alpha}^\ast \sigma^{a\alpha}_{~~\beta} \bar{Z}_1^\beta + \bar{Z}_{2\alpha}^\ast \sigma^{a\alpha}_{~~\beta} \bar{Z}_2^\beta + \bar{Z}_{3\alpha}^\ast \sigma^{a\alpha}_{~~\beta} \bar{Z}_3^\beta + \bar{Z}_{4\alpha}^\ast \sigma^{a\alpha}_{~~\beta} \bar{Z}_4^\beta \,. \label{i10}
\eeq
As the cuprates do not display co-existence between incommensurate SDW and $(\pi, \pi)$ N\'eel ordering, we will only be interested in cases in which (\ref{i10}) vanishes for all components $a$. A solution with only a unidirectional SDW at $(\pi, \pi) \pm 2 \vec{q}_1$ requires that
\beq
\bar{Z}_{2,4}^\alpha = 0 \quad ; \quad \bar{Z}_{3}^\alpha = \varepsilon^{\alpha\beta} \bar{Z}_{1 \beta}^\ast
\label{i11}
\eeq
Then the SDW ordering is described by
\bea
{\bm n} (\vec{r}) &=& \varepsilon_{\alpha\gamma} \bar{Z}_1^\gamma \, {\bm \sigma}^{\alpha}_{~\beta} \bar{Z}_{1}^\beta \, e^{2 i \vec{q}_1 \cdot \vec{r}} + \mbox{c.c.} \nonumber \\
&=& ({\bm m}_1 + i {\bm m}_2) e^{2 i \vec{q}_1 \cdot \vec{r}} + \mbox{c.c.} \,, \label{i12}
\eea
where ${\bm m}_{1,2}$ are real vectors obeying
\beq
{\bm m}_1 \cdot {\bm m}_1 = {\bm m}_2 \cdot {\bm m}_2 = \left(|\bar{Z}_1^\alpha|^2\right)^2 \quad ; \quad {\bm m}_1 \cdot {\bm m}_2 = 0\,. \label{i13}
\eeq
So we see that (\ref{i12}) is a {\it spiral\/} SDW, and not a collinear `stripe' SDW. Such spiral SDW states have been considered in many studies, and recently in Ref.~\cite{Metzner2022}.
But the present approach of condensing $Z^\alpha$ does not lead to unidirectional `stripe' states with collinear spin correlations, of the type observed in recent numerical studies \cite{Shiwei2022,Ferrero2022}.

We now turn to a microscopic mechanism for the condensation of $Z^\alpha$ at an incommensurate wavevector. Our idea is that the coupling of the $Z^\alpha$ to the Fermi pockets of the FL* phase will lead to a $Z^\alpha$ self-energy, and this self-energy will lead to a minimum in the dispersion of the $Z^\alpha$ at $(\pi, \pi) \pm \vec{q}_{1,2}$. A similar approach was used early on in Ref.~\cite{SS93}, but in a theory of the paramagnons $n^a$ coupled to the Fermi pockets. Our point here is that the existence of Fermi pockets which do not have a Luttinger volume requires fractionalization, and so we should carry out the computation using the fractionalized spinon excitations $Z^\alpha$ rather than the paramagnons $n^a$. 

We write the Hamiltonian for the electron layer and the first ancilla layer in Eq. (\ref{a3}) in the form
\begin{equation}
\label{eq:Hcf}
H_{cf} = \sum_{\kk} \Bigg[
\ec{\kk} \cd{\kk,\alpha} c_{\kk}^{\alpha} 
+\ef{\kk} \fd{\kk,\alpha} f_{\kk}^{\alpha} 
+ \bar\Phi \fd{\kk,\alpha} c_{\kk}^{\alpha} 
+ \bar\Phi^{*} \cd{\kk, \alpha} f_{\kk}^{\alpha}  
\Bigg] \,.
\end{equation}
This yields the normal fermion Green's functions
\begin{align}
\label{eq:Gc}
 \gc{\kk}{\iw} &= \frac{\iw - \ef{\kk}}{(\iw-\ec{\kk})(\iw-\ef{\kk}) - |\bar\Phi|^2} 
 = \gcc{\kk}{\iw} \,, \\
 %%%
 \label{eq:Gf}
 \gf{\kk}{\iw} &= \frac{\iw - \ec{\kk}}{(\iw-\ec{\kk})(\iw-\ef{\kk}) - |\bar\Phi|^2} 
 = \gff{\kk}{\iw} \,,
\end{align}
where,
\begin{align}
\label{eq:E12}
\Ea{\kk} &= \frac{\ec{\kk}+\ef{\kk} + \sqrt{(\ec{\kk}-\ef{\kk})^2 + 4|\bar\Phi|^2}}{2} \,,\nonumber \\
\Eb{\kk} &= \frac{\ec{\kk}+\ef{\kk} - \sqrt{(\ec{\kk}-\ef{\kk})^2 + 4|\bar\Phi|^2}}{2} 
\end{align} 
and $\ec{\kk}$, $\ef{\kk}$ are defined in Eqs. (\ref{ecdef}) and (\ref{efdef}). 

The $Z^\alpha$ are described by a CP$^{1}$ field theory,
\begin{equation}
\label{eq:Lz}
\mathcal{L} = \frac{1}{g} |(\partial_{\mu} - ia_{\mu}) Z^{\alpha}|^{2} + i\lambda (|Z^{\alpha}|^2 -1) \,.
\end{equation}
Here $i\lambda= \overline\lambda + i\tilde\lambda$ where $\tilde\lambda$ is the fluctuating part of $\lambda$ and the saddle point $\overline\lambda$ is the Lagrange multiplier imposing the constraint on the $Z$ boson density.
Thus, ignoring the gauge field, the $Z$ boson Green's function is 
\begin{equation}
\label{eq:Gz}
\gz{\kk}{\iw} = \frac{g}{\omega^2 + \ez{\kk} + \overline\lambda - \Pi(\kk,\iw)} \,,
\end{equation} where $\Pi(\kk,\iw)$ denotes the self-energy of the $Z$ boson.
The bare $Z$ boson propagator is given by 
\begin{equation}
\label{eq:Gzzero}
\gzzero{\kk}{\iw} = \gzf{\kk}{\iw} \,,
\end{equation}
where $\Ez{\kk} = \sqrt{\ez{\kk} + \overline\lambda}$, with $\ez{\kk} = k^2$ in the continuum, but on the lattice we use $\ez{\kk} = v(1 - (\cos(k_{x}) + \cos(k_{y}))/2 )$ leading to the form $\ez{\kk} \sim v k^2/4$  near $\kk=0$. 
We write the coupling of $Z$ bosons to the $c$ and $f$ layers in the first terms in Eq. (\ref{a4}) as
\begin{equation}
\label{eq:Hz}
H_{Z} = \sum_{i} \eta_{i} \zd{i\gamma} \sigma^{a\gamma}_{~~\delta} Z_{i}^{\delta} \Big[
-\js \cd{i\alpha} \sigma^{a\alpha}_{\beta} c_{i}^{\beta} 
+\jp \fd{i\alpha} \sigma^{a\alpha}_{~~\beta} f_{i}^{\beta} 
\Big] \,,
\end{equation}
where we recall that $\eta_{i} = (-1)^{x_{i}+y_{i}}$.
In addition, we will also consider the influence of a term of the form,
\begin{equation}
 \label{eq:Hzc}
% H_{zcf}^{a} = \sum_{i} \jc \zd{i\gamma} \za{i\delta} ( \cd{i}\fa{i} + \fd{i}\ca{i} ) 
% ~~~ \text{or} ~~~
 H_{3} = \jc \sum_{i} \eta_i \zd{i\gamma} \sigma^{a\gamma}_{~~\delta} Z_{i}^{\delta} \left[ \cd{i\alpha}\sigma^{a\alpha}_{~~\beta}f_{i}^{\beta} + \fd{i\alpha}\sigma^{a\alpha}_{~~\beta}c_{i}^{\beta} \right] \,.
\end{equation}
This term involves a non-local exchange between the top two layers, and is clearly permitted by the symmetries of the problem.

The coupling terms in Eq. (\ref{eq:Hz}) and Eq. (\ref{eq:Hzc}) contribute to self-energy corrections to the $Z$ boson propagator. At the bare level the $Z$ boson propagator at $\iw=0$, $\gzzero{\kk}{\iw=0}$, has a maximum at $\kk=0$. We are interested in the case where the renormalized $Z$ propagator at $\iw=0$, $\gz{\kk}{\iw=0}$, has a maximum at $\kk \ne 0$  or, equivalently, the inverse $Z$ propagator $\gz{\kk}{\iw=0}^{-1}$ has its minimum at $\kk \ne 0$ . This means that at low temperatures the $Z$ boson can condense at this non-zero $\kk$ leading to a non-trivial spin order. In our case, that may correspond to a spiral or a double spiral. 

%%%%%%%%%%%%%%%%%%%%%%%%%%%%%%%%%%%%%%%%%%%%%%
\subsection{Spinon self-energy}
\label{sec:spinonself}

%%%%%%%%%%%%%%%%%%%%%%%%%%%%%%%
\begin{figure}
 \centering 
 \subfloat[]{\includegraphics[width=0.15\textwidth]{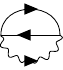}} ~~~~~
 \subfloat[]{\includegraphics[width=0.15\textwidth]{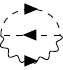}} ~~~~~
 \subfloat[]{\includegraphics[width=0.15\textwidth]{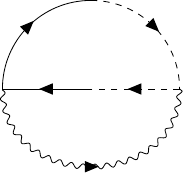}} ~~~~~
 \subfloat[]{\includegraphics[width=0.15\textwidth]{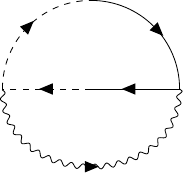}}
 \caption{Self-energy diagrams from the $\js$ and $\jp$ vertices. (a) Diagram from only $\js$ vertex proportional to $\js^2$. (b) Diagram from only $\jp$ vertex proportional to $\jp^2$. (c)-(d) Diagrams combining $\js$ and $\jp$ vertices, which involves the mixed propagators $G_{cf}$ and $G_{fc}$, and is  proportional to $\js \jp$ with an additional suppression by $|\bar\Phi|^2$. Solid lines correspond to $c$ propagators, dashed lines correspond to $f$ propagators, and wiggly lines correspond to bare $Z$ boson propagator. The mixed line with solid and dashed corresponds to $G_{cf}$ and $G_{fc}$. }
 \label{fig:se}
\end{figure}
%%%%%%%%%%%%%%%%%%%%%%%%%%%%%%%

Let us first write the  lowest-order self-energy contribution from the $H_{Z}$ term in Eq. (\ref{eq:Hz}). It has two diagams shown in Fig. \ref{fig:se} (a)-(b) leading to the following forms,
\begin{align}
\label{eq:se_z1}
\Pi_{\ref{fig:se} a}(\kk,\iw) &= - \frac{6\js^{2}}{\beta^{2}} \sum_{\iw_{1},\iw_{2}} \int_{\kk} \gc{\kk_{1}}{\iw_{1}} \gc{\kk_{2}}{\iw_{2}} \gzzero{\kk-\kk_{1}+\kk_{2} -\qp}{\iw+\iw_{2}-\iw_{1}} 
\,, \\
%%%
\label{eq:se_z2}
\Pi_{\ref{fig:se} b}(\kk,\iw)
&= - \frac{6\jp^{2}}{\beta^{2}} \sum_{\iw_{1},\iw_{2}}  \int_{\kk} \gf{\kk_{1}}{\iw_{1}} \gf{\kk_{2}}{\iw_{2}} \gzzero{\kk-\kk_{1}+\kk_{2} -\qp}{\iw+\iw_{2}-\iw_{1}}  \,,
\end{align}
where we have used a short-hand notation,
\begin{equation}
\int_{\kk} = \frac{1}{(2\pi)^4} \int  d\kk_{1} d\kk_{2} \,,
\end{equation}
and $\qp = (\pi,\pi)$, which arises by writing $\eta_{i} = e^{i\qp\cdot\vec{R}_{i}}$.
It turns out that for both these diagrams $-\Pi_{\ref{fig:se} a/b}(\kk,0)$ has its minimum at $\kk=0$, and thus $\gz{\kk}{0}^{-1}_{\ref{fig:se}ab}=\gzzero{\kk}{0}^{-1}-\Pi_{\ref{fig:se} a}(\kk,0)-\Pi_{\ref{fig:se}b}(\kk,0)$ always has its minimum at $\kk=0$.  
In addition to these diagrams, there are also self-energy diagrams involving mixed  propagators, $G_{cf}$ and $G_{fc}$, which lead to a minimum at $\kk \neq 0$ in $\gz{\kk}{0}^{-1}$. The corresponding diagrams are shown in Fig. \ref{fig:se} (c) and (d). However, these diagrams are suppressed by a factor of $|\bar\Phi|^2$ arising in the mixed propagators, which is quite small. So these contributions will be always subdominant compared to the previous diagrams. For completeness we quote their expressions here,
\begin{align}
\label{eq:se_z3}
\Pi_{\ref{fig:se} c}(\kk,\iw) &= \frac{6\js\jp}{\beta^{2}} \sum_{\iw_{1},\iw_{2}} \int_{\kk}  \gcf{\kk_{1}}{\iw_{1}} \gfc{\kk_{2}}{\iw_{2}} \gzzero{\kk-\kk_{1}+\kk_{2} -\qp}{\iw+\iw_{2}-\iw_{1}} 
\,, \\
%%%
\label{eq:se_z4}
\Pi_{\ref{fig:se} d}(\kk,\iw)
&= \frac{6\jp \js}{\beta^{2}} \sum_{\iw_{1},\iw_{2}}  \int_{\kk}  \gfc{\kk_{1}}{\iw_{1}} \gcf{\kk_{2}}{\iw_{2}} \gzzero{\kk-\kk_{1}+\kk_{2} -\qp}{\iw+\iw_{2}-\iw_{1}}  \,.
\end{align}

%%%%%%%%%%%%%%%%%%%%%%%%%%%%%%%
\begin{figure}
 \centering 
 \subfloat[]{\includegraphics[width=0.15\textwidth]{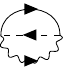}} ~~~~~
 \subfloat[]{\includegraphics[width=0.15\textwidth]{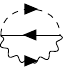}} ~~~~~
 \subfloat[]{\includegraphics[width=0.15\textwidth]{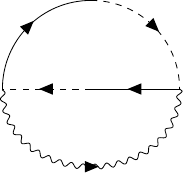}} ~~~~~
 \subfloat[]
{\includegraphics[width=0.15\textwidth]{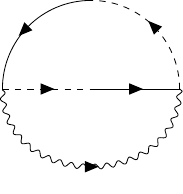}}
 \caption{Self-energy diagrams from the $\jc$ vertex. Solid lines correspond to $c$ propagators, dashed lines correspond to $f$ propagators, and wiggly line correspond to bare $Z$ boson propagators. The mixed lines with solid and dashed correspond to $G_{cf}$ and $G_{fc}$. }
 \label{fig:se2}
\end{figure}
%%%%%%%%%%%%%%%%%%%%%%%%%%%%%%%

\begin{figure}
\begin{center}
\includegraphics[width=\textwidth]{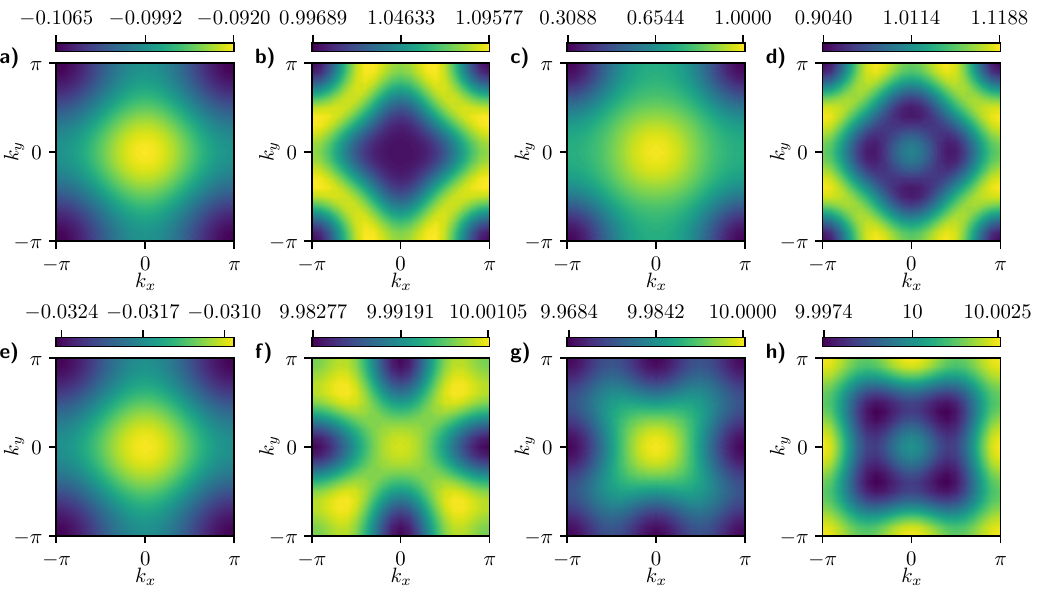}
\end{center}
\caption{Self-energy contributions and bosonic propagators. In subfigures a) and e) the self-energy contributions $\left[-\Pi_{\ref{fig:se2} a}(\kk,0)-\Pi_{\ref{fig:se2} b}(\kk,0)\right]/6\jc^2$ are shown for $m_Z^2=1,$ $\beta=10$ and $m_Z^2=10,$ $\beta=10$, respectively. The remaining subfigures show the inverse propagator of the $Z$ boson $\gz{\kk}{0}^{-1}_{\ref{fig:se2} ab}$, computed according to Eq. (\ref{eq:GzSC}), at $\iw=0$ for different values of $\jc,\,m_Z^2$ and $\beta$ 
(\textbf{b)}: $m_Z^2=1,\beta=10, 6\jc^2=552.5$, 
\textbf{c)}: $m_Z^2=1,\beta=10,6\jc^2=600$, 
\textbf{d)}: $m_Z^2=1,\beta=1000,6\jc^2=378$, 
\textbf{f)}: $m_Z^2=10,\beta=10,6\jc^2=5300$, 
\textbf{g)}: $m_Z^2=10,\beta=10,6\jc^2=5319$, 
\textbf{h)}: $m_Z^2=10,\beta=5,6\jc^2=9655$). These values were chosen to showcase different constellations in which the minimum of the inverse boson propagator can occur at $\kk\neq 0$. For the boson dispersion a value of $v=4$ was used.}
\label{fig:ZProp}
\end{figure}

The self-energy contribution arising from the $H_{3}$ term has two diagrams involving normal propagators, shown in Fig. \ref{fig:se2} (a) and (b), giving the following expressions,
\begin{align}
\label{eq:se_J3_1}
\Pi_{\ref{fig:se2} a}(\kk,\iw) 
&= - \frac{6\jc^{2}}{\beta^{2}} \sum_{\iw_{1},\iw_{2}} \int_{\kk} 
\gc{\kk_{1}}{\iw_{1}} \gf{\kk_{2}}{\iw_{2}} \gzzero{\kk-\kk_{1}+\kk_{2} -\qp}{\iw+\iw_{2}-\iw_{1}}   \,, \\
%%%%%
\label{eq:se_J3_2}
\Pi_{\ref{fig:se2} b}(\kk,\iw) &= - \frac{6\jc^{2}}{\beta^{2}} \sum_{\iw_{1},\iw_{2}} \int_{\kk} 
 \gf{\kk_{1}}{\iw_{1}} \gc{\kk_{2}}{\iw_{2}} \gzzero{\kk-\kk_{1}+\kk_{2} -\qp}{\iw+\iw_{2}-\iw_{1}} 
\,.
%%%
\end{align}
For these diagrams $-\Pi_{\ref{fig:se2} a/b}$ has its minimum at $\kk\neq 0$ (see Fig. \ref{fig:ZProp}) and thus for sufficiently large values of $\jc$ the inverse propagator of the $Z$ boson, $\gz{\kk}{0}^{-1}_{\ref{fig:se2} ab}=\gzzero{\kk}{0}^{-1}-\Pi_{\ref{fig:se2} a}(\kk,0)-\Pi_{\ref{fig:se2}b}(\kk,0)$, has a minimum at $\kk \ne 0$. 
In addition, as before, there are two more diagrams involving mixed propagators, shown in Fig. \ref{fig:se2} (c) and (d), with the following expressions,
\begin{align}
\label{eq:se_J3phi_1}
\Pi_{\ref{fig:se2} c}(\kk,\iw) 
&= - \frac{6\jc^{2}}{\beta^{2}} \sum_{\iw_{1},\iw_{2}} \int_{\kk} 
\gcf{\kk_{1}}{\iw_{1}} \gcf{\kk_{2}}{\iw_{2}} \gzzero{\kk-\kk_{1}+\kk_{2} -\qp}{\iw+\iw_{2}-\iw_{1}}  \,, \\
%%%%%
\label{eq:se_J3phi_2}
\Pi_{\ref{fig:se2} d}(\kk,\iw) &= - \frac{6\jc^{2}}{\beta^{2}} \sum_{\iw_{1},\iw_{2}} \int_{\kk} 
 \gfc{\kk_{1}}{\iw_{1}} \gfc{\kk_{2}}{\iw_{2}} \gzzero{\kk-\kk_{1}+\kk_{2} -\qp}{\iw+\iw_{2}-\iw_{1}} 
\,.
%%%
\end{align}
These are again suppressed by an additional small factor of $|\bar\Phi|^2$ and so these are subdominant compared to the previous diagrams. In Appendix \ref{sec:se_exp} we provide detailed expressions of the self-energy terms. 

To make the renormalization of the $Z$ boson partially self-consistent we employ a similar approach as Ref.~\cite{SS93}. To this end, we rewrite its propagator defined in Eq. (\ref{eq:Gz}) as
\begin{equation}
\label{eq:GzSC}
\gz{\kk}{\iw} = \frac{g}{\omega^2 + \ez{\kk} + m_Z^2 - \Pi(\kk,\iw)+ \Pi(0,0)} \,,
\end{equation}
where $m_Z^2= \overline\lambda - \Pi(0,0)$ is the boson mass gap. Thus, to take into account the renormalized mass gap, in Eq. (\ref{eq:Gzzero}) we replace $\overline\lambda$ by $m_Z^2$, effectively using  $\gzzero{\kk}{\iw} =(\omega^2 + \ez{\kk} + m_Z^2)^{-1}$ in Eqs. (\ref{eq:se_z1})-(\ref{eq:se_z2}) and (\ref{eq:se_z3})-(\ref{eq:se_J3phi_2}). As a result, the closing of the bosonic mass gap due to the renormalization process affects the renormalization of the $Z$ boson.

In Fig. \ref{fig:ZProp} we show self-energy contributions $\left[-\Pi_{\ref{fig:se2} a}(\kk,0)-\Pi_{\ref{fig:se2} b}(\kk,0)\right]$ (in units of $6\jc^2$) along with the inverse propagator of the $Z$ boson resulting from these corrections. As it can be seen, the self-energy contributions (multiplied by a factor of $-1$) are negative and feature a minimum at $\kk= \qp$. Thus, for a sufficiently large value of $6 \jc^2$ the $Z$  boson inverse propagator, computed according to Eq. (\ref{eq:GzSC}), features a minimum at $\kk\neq 0$. For smaller values of $\jc$ its minimum remains at $\kk=0$, while for intermediate values its minimum can behave in one of several ways before eventually reaching $\kk=\qp$ for sufficiently large values of $\jc$. For these intermediate values of $\jc$ we have observed instances where the  minimum  moved along the diagonal ($k_x=k_y$) or the vertical/horizontal ($k_y=0/k_x=0$). Additionally, it has jumped from $\kk=0$ via $\kk=(\pi,0)$ to $\kk=\qp$ or even directly. Several of such possibilities are illustrated in Fig. \ref{fig:ZProp}, where we show bosonic propagators with degenerate minima between $0$ and $\qp$ (\textbf{b)}) or $(\pi,0)$ and $(\pi,\pi)$ (\textbf{g)}) as well as propagators with minima at $\qp$ (\textbf{c)}), minima at $(k,0),(0,k)$ (\textbf{d)}), minima at $(\pi,0),(0,\pi)$ (\textbf{f)}) and minima at $(k,k)$ (\textbf{h)}), where $k\in [0,\pi]$ and we have only listed minima in the upper right quadrant.

%%%%%%%%%%%%%%%%%%%%%%%%%%%%%%%%%%%%%%%%%%%%%%%%%%%%%%%%%%%%%%%%%

\subsection{Fermi surfaces}

This section will briefly present the Fermi surfaces induced by incommensurate SDW order across the transition from the FL* metal. Along with the spiral SDW obtained in (\ref{i12}), we will also consider collinear SDW states for completeness in the following subsections. The latter states are `stripes' because they have co-existing charge density wave order.  

\subsubsection{Collinear bi-directional SDW}

Collinear SDWs are characterized by complex order parameters $\Phi_{x}^a$ and $\Phi_{y}^a$, taking the place of the real vectors $m^a_{1,2}$ in (\ref{i12}) for spiral SDWs. These determine the spin density via
\beq
S^a ({\bf r}) = \mbox{Re} \left[ e^{i \vec{K}_x \cdot {\bm r}} \Phi_{x}^a ({\bm r})  +  e^{i \vec{K}_y \cdot {\bm r}} \Phi_{y}^a ({\bm r})\right] \label{st1}
\eeq
where
\beq
\vec{K}_x = (3 \pi/4, \pi) \quad, \quad \vec{K}_y = (\pi, 3 \pi/4)\,. \label{eq:wave}
\eeq
An order parameter with $\mbox{arg} (\Phi_{xa}) = (0, \pi/4, \pi/2, 3\pi/4, \pi, 5 \pi/4, 3\pi/2, 7\pi/4) $ is bond-centered, and one with $\mbox{arg} (\Phi_{xa}) = (\pi/8, 3\pi/8, 5\pi/8, 7\pi/8, 9\pi/8, 11 \pi/8, 13\pi/8, 15\pi/8) $ is site-centered. We chose a commensurate wavevector to obtain a smaller unit cell.

The mean-field Hamiltonian for the bidirectional SDW will be given by $H_{cf}^a+H_{cf}^c$, with the wave vectors in  (\ref{eq:wave}). In momentum space $H_{cf}^a$ is given by equation (\ref{eq:Hcf}) and $H_{cf}^c$ looks as follows:
\begin{equation}
    H_{cf}^c=\sum_{\vec{k}}-J_s \left(  \Phi_x c^{\dag}_{\vec{k}+\vec{K}_x,\alpha} \, \sigma^{z \alpha}_{~~\beta} \, c_{\vec{k}}^{\beta}+ \Phi_y c^{\dag}_{\vec{k}+\vec{K}_y,\alpha} \, \sigma^{z \alpha}_{~~\beta} \, c_{\vec{k}}^{\beta} \right) +h.c.+J_\perp(...)+J_3(...)\,,
    \label{eq:H_SDW}
\end{equation}
where the $J_\perp$ term is the same as $J_s$ term with $c \rightarrow f$. We also include $J_3$ term which couples $c$ and $f$ fermions in the same way, see (\ref{eq:Hzc}).

\begin{figure}[h]
\begin{minipage}[h]{0.45\linewidth}
  \center{\includegraphics[width=1\linewidth]{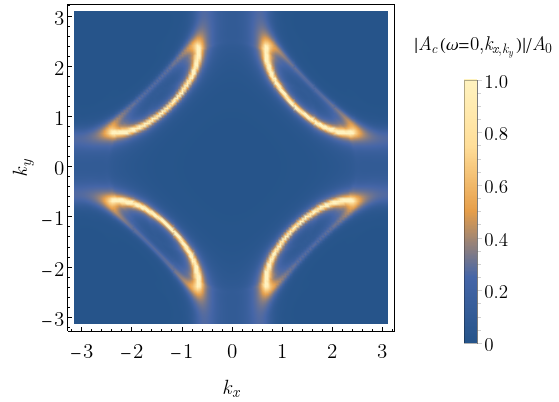}}
  \\$\Phi_s=0.01$
  \end{minipage} 
\hfill
  \begin{minipage}[h]{0.45\linewidth}
  \center{\includegraphics[width=1.\linewidth]{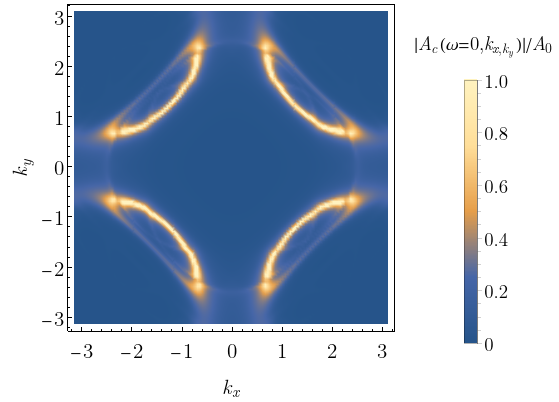}}
  \\$\Phi_s=0.02$
  \end{minipage} 
\caption{Spectral weight in the collinear bi-directional SDW. $J_s=2J_\perp=1$, $J_3=0.2$}
\label{fig:fermi_surface_stripe_1}
\end{figure}

\begin{figure}[h]
\begin{minipage}[h]{0.45\linewidth}
  \center{\includegraphics[width=1\linewidth]{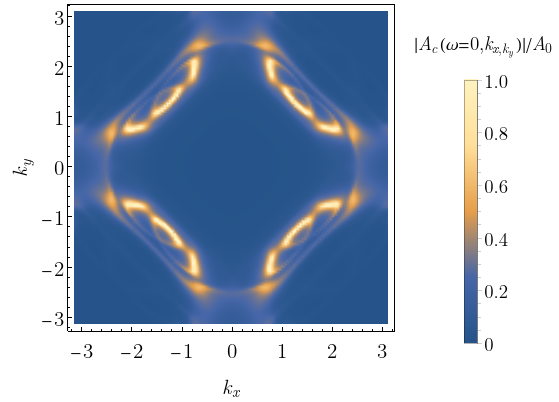}}
  \\$\Phi_s=0.04$
  \end{minipage} 
\hfill
  \begin{minipage}[h]{0.45\linewidth}
  \center{\includegraphics[width=1.\linewidth]{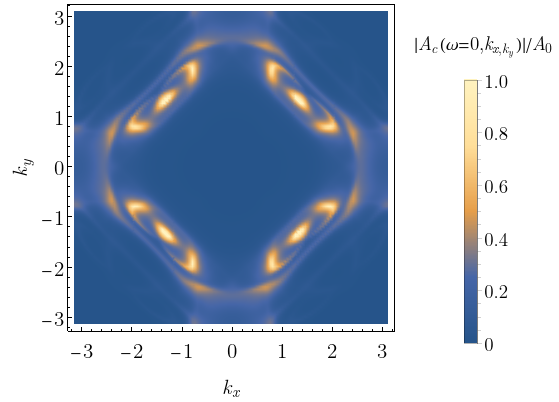}}
  \\$\Phi_s=0.06$
  \end{minipage} 
\caption{Spectral weight in the collinear bi-directional SDW. $J_s=2J_\perp=1$, $J_3=0.2$}
\label{fig:fermi_surface_stripe_2}
\end{figure}

The Hamiltonian can be diagonalized in the following basis of 128 elements given by
$\Psi_k=(c_{k_x+Q_i,k_y+Q_j},f_{k_x+Q_i,k_y+Q_j})$, where $Q_i=(0,3 \pi/4,6 \pi/4,9 \pi/4,\pi,7 \pi/4,2 \pi/4,5 \pi/4)$.  We can compute a spectral weight after the SDW order emerges. For simplicity we put $J_s=2J_\perp=1$, $J_3=0$ and change only $\Phi_x=\Phi_s$, $\Phi_y=\Phi_s e^{i \pi /8}$. Chemical potentials and hybridization were chosen to be as in Fig. \ref{fig:fermi_surface_1}. Figure \ref{fig:fermi_surface_stripe_1}, \ref{fig:fermi_surface_stripe_2} shows that hole pockets evolve into more complicated Fermi surfaces with the possible gap closing in the anti-nodal region.

% For different parameters $J_s=2J_\perp=1$, see Figure \ref{fig:fermi_surface_stripe_3}, \ref{fig:fermi_surface_stripe_4} the evolution of hole pockets is different and we find it hard to relate to an existing experiments. 

%      \begin{figure}[h]
% \begin{minipage}[h]{0.45\linewidth}
%   \center{\includegraphics[width=1\linewidth]{spectral_weight_c_phi_1=0.02.png}}
%   \\$\Phi_s=0.02$
%   \end{minipage} 
% \hfill
%   \begin{minipage}[h]{0.45\linewidth}
%   \center{\includegraphics[width=1.\linewidth]{spectral_weight_c_phi_1=0.04.png}}
%   \\$\Phi_s=0.04$
%   \end{minipage} 
% \caption{Spectral weight in the striped phase. $J_s=2J_\perp=1$, $J_3=0$}
% \label{fig:fermi_surface_stripe_3}
% \end{figure}

%      \begin{figure}[H]
% \begin{minipage}[h]{0.45\linewidth}
%   \center{\includegraphics[width=1\linewidth]{spectral_weight_c_phi_1=0.06.png}}
%   \\$\Phi_s=0.06$
%   \end{minipage} 
% \hfill
%   \begin{minipage}[h]{0.45\linewidth}
%   \center{\includegraphics[width=1.\linewidth]{spectral_weight_c_phi_1=0.08.png}}
%   \\$\Phi_s=0.08$
%   \end{minipage} 
% \caption{Spectral weight in the striped phase. $J_s=2J_\perp=1$, $J_3=0$}
% \label{fig:fermi_surface_stripe_4}
% \end{figure}

\subsubsection{Collinear unidirectional SDW}

\begin{figure}[h]
\begin{minipage}[h]{0.45\linewidth}
  \center{\includegraphics[width=1\linewidth]{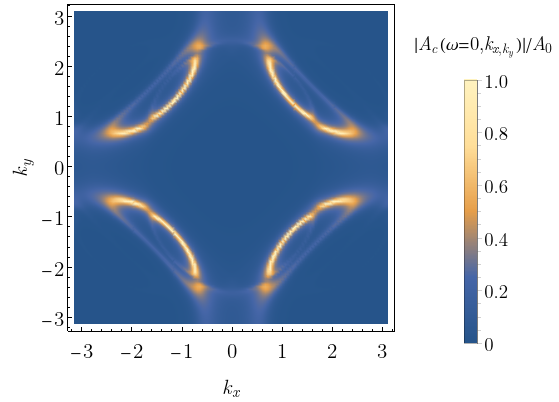}}
  \\$\Phi_s=0.03$
  \end{minipage} 
\hfill
  \begin{minipage}[h]{0.45\linewidth}
  \center{\includegraphics[width=1.\linewidth]{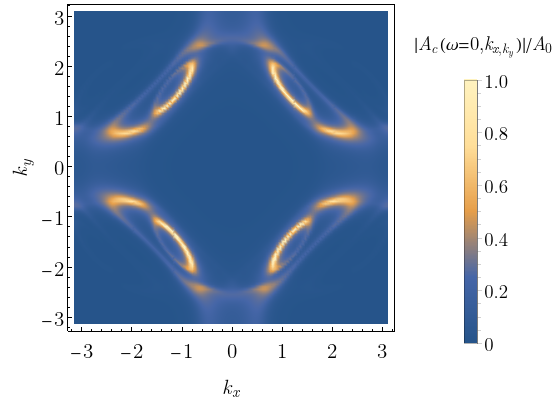}}
  \\$\Phi_s=0.05$
  \end{minipage} 
\caption{Spectral weight in the unidirectional collinear SDW phase. $J_s=2J_\perp=1$, $J_3=0.2$}
\label{fig:fermi_surface_collinear_1}
\end{figure}

Now we consider the case of unidirectional SDW with wave vector $\vec{K}_x = (3 \pi/4, \pi)$. The Hamiltonian will be the same as in a previous case with $\Phi_y=0$. The evolution of the spectral weight is depicted in Fig. \ref{fig:fermi_surface_collinear_1}. We see that as we move deeper into the SDW phase the hole pocket of a different size arises in the nodal region followed by the emergence of the fermion pocket in the anti-nodal region.

\subsection{Spiral unidirectional SDW}

\begin{figure}[h]
\begin{minipage}[h]{0.45\linewidth}
  \center{\includegraphics[width=1\linewidth]{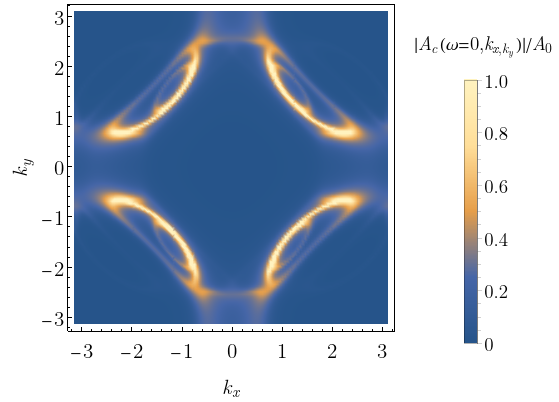}}
  \\$\vec{K}_x=(3\pi/4, \pi)$
  \end{minipage} 
\hfill
  \begin{minipage}[h]{0.45\linewidth}
  \center{\includegraphics[width=1.\linewidth]{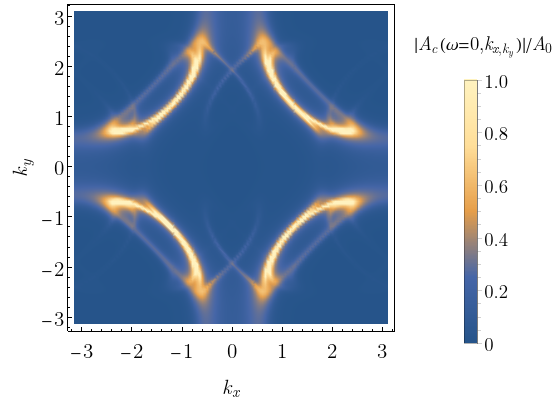}}
  \\$\vec{K}_x=(3\pi/8, \pi)$
  \end{minipage} 
\caption{Spectral weight in the unidirectional spiral SDW phase. $J_s=2J_\perp=1$, $J_3=0.2$, $\Phi_s=0.06$}
\label{fig:fermi_surface_spiral_1}
\end{figure}

We consider spiral SDW with $\vec{S}(r)=\Phi_x(\vec{x}+i\vec{y}) e^{i \vec{K_x} \vec{r}} + (c.c.)$. $H_{cf}^a$ part of the Hamiltonian is not changed while $ H_{cf}^c$ part looks as follows:
\begin{equation}
    H_{cf}^c=-J_s\left( \Phi_x c^{\dag}_{\vec{k}+\vec{K}_x,\alpha} (\sigma^+)^{\alpha}_{~\beta} c_{\vec{k}}^{\beta}+\Phi_x c^{\dag}_{\vec{k},\alpha} (\sigma^-)^{\alpha}_{\beta} c_{\vec{k}+\vec{K}_x}^{\beta} \right) +J_\perp(...)+J_3(...)
    \label{eq:H_SDW_spiral}
\end{equation}
The Hamiltonian can be diagonalized in the following basis:
\begin{equation}
\Psi_{\vec{k}}=(c_{\vec{k},\uparrow},c_{\vec{k},\downarrow},c_{\vec{k}+\vec{K}_x,\uparrow},c_{\vec{k}-\vec{K}_x,\downarrow},f_{\vec{k},\uparrow},f_{\vec{k},\downarrow},f_{\vec{k}+\vec{K}_x,\uparrow},f_{\vec{k}-\vec{K}_x,\downarrow})    
\end{equation}
It is important to stress that even for incommensurate $\vec{K}_x$ the size of the basis is not changed; that is because fermions with spin up can only scatter by wave vector $+\vec{K}_x$ to fermions with spin down, but can not scatter back with wave vector $-\vec{K}_x$. So the present computation can be carried out for arbitrary $\vec{K}$, and the results for commensurate and incommensurate wavevectors are not different. The distribution of the spectral weight, see Figure \ref{fig:fermi_surface_spiral_1}, is similar to a collinear scenario, but the important difference is that original hole pockets do not disappear and reconstruction of the Fermi surface happens on top of the hole pockets. The exact way the Fermi surface is reconstructed would depend on the wave vector $\vec{K}_x$.

\section{Discussion}

An earlier paper \cite{Mascot22} showed that the observed photoemission spectrum in the pseudogap metal of the underdoped hole-doped cuprates \cite{chen2019incoherent,He2011} could be well described by a paramagnon fractionalization theory summarized in Figs.~\ref{fig:ancilla1} and \ref{fig:parafrac}. A similar connection to the experimental data has also been made in the YRZ framework \cite{YRZ_rev}. We view this agreement as evidence in favor of the presence of spin fractionalization in the underdoped cuprates at intermediate temperatures. 

Also notable are recent experimental studies  \cite{Kondo2020} of the very lightly doped state with long-range N\'eel order at wavevector $(\pi, \pi)$. Convincing evidence has recently been obtained for the presence of hole pockets in such a metallic state. 

This situation provided motivation for the studies presented in this paper, as summarized in Fig.~\ref{fig:parent}. We view the pseudogap metal at intermediate temperatures and under-doping as the `parent' of the phase diagram. We described this pseudogap as a metal with hole pockets whose enclosed volume does not equal the Luttinger volume. Consequently, this phase must have fractionalized spin excitations {\it i.e.\/} it is an FL* metal.

We described how the FL* state of the pseudogap metal evolved into a metallic $(\pi, \pi)$ N\'eel state without fractionalization with {\it decreasing\/} doping, as shown by arrow $\mathbb{A}$ in Fig.~\ref{fig:parent}. We used charge neutral bosonic spinons to represent the fractionalized paramagnons, and condensation of such spinons led to the N\'eel state in which all emergent gauge fields were higgsed. 
We described the evolution of the hole pockets across the transition from the FL* state to the metallic N\'eel state in Section~\ref{flstar-sdw}. We are presenting these results as predictions for future observations which are able to follow the Fermi surfaces from the pseudogap metal to the ordered N\'eel state; in particular, Figs.~\ref{fig:fermi_surface_1} and \ref{fig:fermi_surface_2} show how the effective size of the hole pocket doubles \cite{Kaul07} in the full Brillouin zone across the transition from the pseudogap metal to the N\'eel metal. Furthermore, we propose that, in sufficiently clean samples, the evidence for hole pockets in the pseudogap metal state {\it without\/} magnetic order will become sharper than that presented in Ref.~\cite{PDJ11}: this would then constitute direct evidence for an FL* metal with fractionalized excitations. 

We also considered the situation with {\it increasing\/} 
doping from the pseudogap metal, as shown by arrow $\mathbb{B}$ in Fig.~\ref{fig:parent}.
In this direction, the FL* state evolves into a conventional Fermi liquid via an intermediate metallic state with ghost Fermi surfaces \cite{Zhang2020,Zhang2021,nikolaenko2021}.  We studied this route to the confinement of fractionalized excitations, and our results are summarized in the phase diagram in Fig.~\ref{fig:pd}. All the phases in this diagram are described in terms of a gauge theory of the fields collected in Table~\ref{tab1}. The paramagnon field ${\bm n}$ is gauge neutral, and it is related to the fractionalized fields via (\ref{i7}); the paramagnon becomes an elementary excitation in the Fermi liquid phases where all gauge charges are confined.  The effective action for these fields can be deduced from the symmetries listed in Table~\ref{tab1}, and the potential for the bosonic fields appeared in (\ref{a1}). The nature of the transition out of the pseudogap metal to the Fermi liquid, {\it i.e.\/} that between phases A and D in Fig.~\ref{fig:pd}, is the same as that considered in earlier work \cite{Zhang2020,Zhang2021}, and involves critical fluctuations of the Higgs field $\Phi$ coupled to the $c$ and $f$ Fermi surfaces. The second ancilla layer of spins is not important to this critical theory, and so the use of bosonic spinons here, in contrast to the fermionic spinons in the earlier work \cite{Zhang2020,Zhang2021}, does not make a substantial difference. Upon adding spatial disorder to the Yukawa coupling of the Higgs field $\Phi$, such a theory will yield a strange metal with linear-$T$ resistivity in the critical region \cite{Aldape20,Patel:2022gdh}.

Finally, in Section~\ref{sec:stripes} we considered the fate of the FL* metal upon lowering $T$ within the pseudogap, as shown by arrow $\mathbb{C}$ in Fig.~\ref{fig:parent}. Here, we face the important question of whether the fractionalization will survive at lower $T$, or even at $T=0$. There has been no direct experimental evidence in favor of fractionalization at low $T$ so far. Recent numerical studies on the doped Hubbard model \cite{Ferrero2022,Shiwei2022} indicate that the $T=0$ state is 
a conventional striped state with both spin and charge density wave orders. 
In light of this, Section~\ref{sec:stripes} considered the appearance of a confining incommensurate SDW state via the condensation of the bosonic spinons at an incommensurate wavevector. We showed in Section~\ref{sec:spinonself} that coupling of the bosonic spinons to the hole pockets could induce a self-energy so that the spinon dispersion minimum was at an incommensurate wavevector. 
This mechanism for the appearance of incommensurate SDW from hole pockets is similar to that considered in Ref.~\cite{SS93}, although that analysis was expressed in terms of the paramagnon dispersion. We presented predictions for the photoemission spectrum as it evolved from the pseudogap to incommensurate SDW states.

A significant result of our analysis, appearing in Eqs.~(\ref{i7a}-\ref{i13}), is that our spinon condensation mechanism did not induce a collinear SDW with 
co-existing charge stripe order. Instead, the SDWs found in Section~\ref{sec:stripes} carried spiral spin correlations. The basic reasoning is independent of the form of the spinon free energy. These equations show that incommensurate spinons induce an incommensurate SDW {\it along with\/} a commensurate SDW. As such a co-existence is not observed, we impose the requirement that the commensurate SDW vanish: we then find that this can only be achieved by an incommensurate spiral SDW.

Arrow $\mathbb{C}$ in Fig.~\ref{fig:parent} also indicates transitions from FL* to $d$-wave superconductivity and charge density wave. Such transitions were discussed in Refs.~\cite{Shubhayu16a,Shubhayu16b} using fermionic spinons for cases where the parent was a $\mathbb{Z}_2$ spin liquid. The dual fermionic spinon description of the confining instabilities of a parent $\mathbb{CP}^1$ U(1) spin liquid is presented in a companion paper \cite{Christos:2023oru}; the dual of the $\mathbb{CP}^1$ U(1) spin liquid has fermionic spinons moving in $\pi$ flux coupled to an emergent SU(2) gauge field \cite{DQCP3}.

\subsection*{Acknowledgements}

We thank Seamus Davis, Michele Fabrizio, Antoine Georges, Masatoshi Imada, Takeshi Kondo, Peter Johnson, Zhi-Xun Shen, Louis Taillefer, and Ya-Hui Zhang for valuable discussions. 
This research was supported by the U.S. National Science Foundation grant No. DMR-2002850 and by the Simons Collaboration on Ultra-Quantum Matter which is a grant from the Simons Foundation (651440, S.S.). 
J.~v.~M. is supported by a fellowship of the International Max Planck Research School for Quantum Science and Technology (IMPRS-QST).

\appendix 

\section{Self-energy expression}
\label{sec:se_exp}

We can perform the frequency summation in the self-energy expression in Eq. (\ref{eq:se_z1}), and obtain,
\begingroup
\allowdisplaybreaks
\begin{align}
&\Pi_{\ref{fig:se} a}(\kk, \iw) = \frac{6\js^{2}}{(2\pi)^4} \int d\kk_{1} d\kk_{2} d\kk_{3} ~\delta(\kk_{1}-\kk_{2}+\kk_{3}-\kk+\qp) \sum_{i=1}^{12} S_{i}  \,, \\
%%%%%%%%
\label{eq:s1}
&S_{1} = \frac{\nf{\Ea{\kk_{1}}} \nf{\Ea{\kk_{2}}} (\Ea{\kk_{1}} - \ef{\kk_{1}}) (\Ea{\kk_{2}}-\ef{\kk_{2}}) }{(\iw-\Ea{\kk_{1}} +\Ea{\kk_{2}} + \Ez{\kk_{3}} ) (\iw-\Ea{\kk_{1}} +\Ea{\kk_{2}} - \Ez{\kk_{3}} )} \frac{1}{(\Ea{\kk_{1}} - \Eb{\kk_{1}})} \frac{1}{( \Ea{\kk_{2}} -\Eb{\kk_{2}})} \,, \\
%%%%%%%%
\label{eq:s2}
&S_{2} = \frac{\nf{\Ea{\kk_{1}}} \nf{\Eb{\kk_{2}}} (\Ea{\kk_{1}} - \ef{\kk_{1}}) (\Eb{\kk_{2}} - \ef{\kk_{2}}) }{(\iw-\Ea{\kk_{1}} +\Eb{\kk_{2}} + \Ez{\kk_{3}} ) (\iw-\Ea{\kk_{1}} +\Eb{\kk_{2}} - \Ez{\kk_{3}} )} \frac{1}{(\Ea{\kk_{1}} - \Eb{\kk_{1}})} \frac{1}{( \Eb{\kk_{2}} -\Ea{\kk_{2}})} \,, \\
%%%%%%%%
\label{eq:s3}
&S_{3} = \frac{\nf{\Eb{\kk_{1}}} \nf{\Ea{\kk_{2}}} (\Eb{\kk_{1}} - \ef{\kk_{1}}) (\Ea{\kk_{2}} -\ef{\kk_{2}}) }{(\iw-\Eb{\kk_{1}} +\Ea{\kk_{2}} + \Ez{\kk_{3}} ) (\iw-\Eb{\kk_{1}} +\Ea{\kk_{2}} - \Ez{\kk_{3}} )} \frac{1}{(\Eb{\kk_{1}} - \Ea{\kk_{1}})} \frac{1}{( \Ea{\kk_{2}} -\Eb{\kk_{2}})} \,, \\ 
%%%%%%%%
\label{eq:s4} 
&S_{4} = \frac{\nf{\Eb{\kk_{1}}} \nf{\Eb{\kk_{2}}} (\Eb{\kk_{1}} - \ef{\kk_{1}}) (\Eb{\kk_{2}} -\ef{\kk_{2}}) }{(\iw-\Eb{\kk_{1}} +\Eb{\kk_{2}} + \Ez{\kk_{3}} ) (\iw-\Eb{\kk_{1}} +\Eb{\kk_{2}} - \Ez{\kk_{3}} )} \frac{1}{(\Eb{\kk_{1}} - \Ea{\kk_{1}})} \frac{1}{( \Eb{\kk_{2}} -\Ea{\kk_{2}})} \,, \\
%%%%%%%%
\label{eq:s5}
&S_{5} = - \frac{(\Ea{\kk_{2}} -\ef{\kk_{2}})}{2\Ez{\kk_{3}} ( \Ea{\kk_{2}} -\Eb{\kk_{2}})} 
\frac{\nf{\Ea{\kk_{2}}} \nb{\Ez{\kk_{3}}} (\iw - \ef{\kk_{1}} + \Ea{\kk_{2}} +\Ez{\kk_{3}}) }{(\iw-\Ea{\kk_{1}} +\Ea{\kk_{2}} + \Ez{\kk_{3}} ) (\iw-\Eb{\kk_{1}} +\Ea{\kk_{2}} + \Ez{\kk_{3}} )}  \,, \\
%%%%%%%%
\label{eq:s6}
&S_{6} = - \frac{( \Eb{\kk_{2}} -\ef{\kk_{2}})}{2\Ez{\kk_{3}} ( \Eb{\kk_{2}} -\Ea{\kk_{2}})} 
\frac{\nf{\Eb{\kk_{2}}} \nb{\Ez{\kk_{3}}} (\iw - \ef{\kk_{1}} + \Eb{\kk_{2}} +\Ez{\kk_{3}}) }{(\iw-\Ea{\kk_{1}} +\Eb{\kk_{2}} + \Ez{\kk_{3}} ) (\iw-\Eb{\kk_{1}} +\Eb{\kk_{2}} + \Ez{\kk_{3}} )}  \,, \\
%%%%%%%%
\label{eq:s7} 
&S_{7} = \frac{(\Ea{\kk_{2}} -\ef{\kk_{2}})}{2\Ez{\kk_{3}} (\Ea{\kk_{2}} -\Eb{\kk_{2}})} 
\frac{\nf{\Ea{\kk_{2}}} \nb{-\Ez{\kk_{3}}} (\iw - \ef{\kk_{1}} + \Ea{\kk_{2}} -\Ez{\kk_{3}}) }{(\iw-\Ea{\kk_{1}} +\Ea{\kk_{2}} - \Ez{\kk_{3}} ) (\iw-\Eb{\kk_{1}} +\Ea{\kk_{2}} - \Ez{\kk_{3}} )}  \,, \\
%%%%%%%%
\label{eq:s8} 
&S_{8} = \frac{(\Eb{\kk_{2}} -\ef{\kk_{2}})}{2\Ez{\kk_{3}} ( \Eb{\kk_{2}} -\Ea{\kk_{2}})} 
\frac{\nf{\Eb{\kk_{2}}} \nb{-\Ez{\kk_{3}}} (\iw - \ef{\kk_{1}} + \Eb{\kk_{2}} -\Ez{\kk_{3}}) }{(\iw-\Ea{\kk_{1}} +\Eb{\kk_{2}} - \Ez{\kk_{3}} ) (\iw-\Eb{\kk_{1}} +\Eb{\kk_{2}} - \Ez{\kk_{3}} )}  \,, \\
%%%%%%%%
\label{eq:s9} 
&S_{9} = \frac{(\Ea{\kk_{1}} -\ef{\kk_{1}})}{2\Ez{\kk_{3}} (\Ea{\kk_{1}} - \Eb{\kk_{1}})} 
\frac{\nf{\Ea{\kk_{1}}-\Ez{\kk_{3}}} [\nf{\Ea{\kk_{1}}} + \nb{\Ez{\kk_{3}}}] (\iw - \Ea{\kk_{1}} + \ef{\kk_{2}} +\Ez{\kk_{3}}) }{(\iw-\Ea{\kk_{1}} +\Ea{\kk_{2}} + \Ez{\kk_{3}} ) (\iw-\Ea{\kk_{1}} +\Eb{\kk_{2}} + \Ez{\kk_{3}} )}  \,, \\
%%%%%%%%
\label{eq:s10} 
&S_{10} = - 
\frac{\nf{\Ea{\kk_{1}}+\Ez{\kk_{3}}} [\nf{\Ea{\kk_{1}}} + \nb{-\Ez{\kk_{3}}}] (\iw - \Ea{\kk_{1}} + \ef{\kk_{2}} -\Ez{\kk_{3}}) (\Ea{\kk_{1}} -\ef{\kk_{1}}) }{2\Ez{\kk_{3}} (\Ea{\kk_{1}} - \Eb{\kk_{1}})(\iw-\Ea{\kk_{1}} +\Ea{\kk_{2}} - \Ez{\kk_{3}} ) (\iw-\Ea{\kk_{1}} +\Eb{\kk_{2}} - \Ez{\kk_{3}} )}  \,, \\
%%%%%%%%
\label{eq:s11}
&S_{11} = \frac{\nf{\Eb{\kk_{1}}-\Ez{\kk_{3}}} [\nf{\Eb{\kk_{1}}} + \nb{\Ez{\kk_{3}}}] (\iw - \Eb{\kk_{1}} + \ef{\kk_{2}} +\Ez{\kk_{3}}) (\Eb{\kk_{1}} -\ef{\kk_{1}}) }{2\Ez{\kk_{3}} (\Eb{\kk_{1}} - \Ea{\kk_{1}})(\iw-\Eb{\kk_{1}} +\Ea{\kk_{2}} + \Ez{\kk_{3}} ) (\iw-\Eb{\kk_{1}} +\Eb{\kk_{2}} + \Ez{\kk_{3}} )}  \,, \\
%%%%%%%%
\label{eq:s12} 
&S_{12} = - 
\frac{\nf{\Eb{\kk_{1}}+\Ez{\kk_{3}}} [\nf{\Eb{\kk_{1}}} + \nb{-\Ez{\kk_{3}}}] (\iw - \Eb{\kk_{1}} + \ef{\kk_{2}} -\Ez{\kk_{3}}) (\Eb{\kk_{1}} -\ef{\kk_{1}}) }{ 2\Ez{\kk_{3}} (\Eb{\kk_{1}} - \Ea{\kk_{1}}) (\iw-\Eb{\kk_{1}} +\Ea{\kk_{2}} - \Ez{\kk_{3}} ) (\iw-\Eb{\kk_{1}} +\Eb{\kk_{2}} - \Ez{\kk_{3}} )}  \,.
\end{align}
\endgroup
The expression for the self-energy $\Pi_{\ref{fig:se} b}$ in Eq. (\ref{eq:se_z2}) is obtained by replacing $\ef{} \to \ec{}$ and $\js\to\jp$ in the above expressions.
The expression for the self-energy $\Pi_{\ref{fig:se2} a}$ in Eq. (\ref{eq:se_J3_1}) is obtained by replacing $\ef{\kk_{2}} \to \ec{\kk_{2}}$ and $\js\to\jc$ in the above expressions, while that for $\Pi_{\ref{fig:se2} b}$ in Eq. (\ref{eq:se_J3_2}) is obtained by replacing $\ef{\kk_{1}} \to \ec{\kk_{1}}$ and $\js\to\jc$. 
The expressions for the self-energies $\Pi_{\ref{fig:se} c},\Pi_{\ref{fig:se} d} $ in Eqs. (\ref{eq:se_z3}) and (\ref{eq:se_z4}) are obtained by replacing all terms in the numerator not involving the   Bose- or Fermi-distribution functions by $|\bar\Phi|^2$, and replacing $\js^2\to \js\jp$ in the above expressions. The expressions for $\Pi_{\ref{fig:se2} c},\Pi_{\ref{fig:se2} d}$ in Eqs. (\ref{eq:se_J3phi_1})  and (\ref{eq:se_J3phi_1}) are obtained analogously,  replacing $\js\to\jc$. 

\section{Pseudogap as a holon metal}

The body of the paper has considered the pseudogap as the non-zero $T$ realization of the FL* state, with hole pockets of spin-1/2, charge $+e$ quasiparticles. An alternative model \cite{sdw09,DCSS15b,DCSS15,WuScheurer1,Scheurer:2017jcp,Sachdev:2018ddg,HeScheurer19,WuScheurer2,Bonetti22} is that of a non-zero $T$ realization of the holon metal, with Fermi pockets of spin-0, charge $+e$ quasiparticles. While the FL* and holon metals are distinct quantum states at $T=0$, their photoemission properties at intermediate temperature can be similar, and there are reasonable comparisons of the holon metal theory to experimental data \cite{HeScheurer19} and numerical studies \cite{WuScheurer1,Scheurer:2017jcp,WuScheurer2}. Moreover, the angle-dependent magnetoresistance (ADMR) observations \cite{Ramshaw22} are insensitive to the spin of the quasiparticles, and so are equally compatible with FL* and the holon metal.

In this appendix, we recall the quantum numbers of the fields in the holon metal approach, and relate them to the fields of the present paper. We show the holon metal fields in Table~\ref{tab3}, using the notation of Ref.~\cite{Sachdev:2018ddg}, along with an additional tilde to distinguish them from those used in the FL* approach here.
\begin{table}[h]
    \centering
    \begin{tabular}{|c|c|c|c|c|c|c|}
\hline
Field & Statistics & U(1)$_1$ & U(1)$_2$ & SU(2)$_S$ & U(1)$_g$ & SU(2)$_g$  \\
\hline 
\hline
$\tilde{f}^p$ chargon & fermion & 0 & 0 & ${\bm 2}$ & 1 & ${\bm 1}$ \\
$\tilde{R}^{\alpha}_{p}$ spinon & boson & 0 & 0 &  $\bar{\bm 2}$ & 0 & ${\bm 2}$ \\
$\tilde{H}^\ell$ Higgs & boson & 0 & 0 &${\bm 3}$ & 0 & ${\bm 1}$ \\
\hline
\end{tabular}
    \caption{As in Table~\ref{tab1}, for fields in the theory of holon metals, in the notation of Ref.~\cite{Sachdev:2018ddg}.}
    \label{tab3}
\end{table}
Comparing the quantum numbers in Table~\ref{tab3} with those in Table~\ref{tab1} and~\ref{tab2}, we can deduce the following relations between the fields
\begin{align}
\tilde{f}^p &= c^\alpha w_{\alpha}^\ast Z^p \nonumber \\
\tilde{R}^{\alpha}_{p} & = w^\alpha Z_p^\ast \nonumber \\
\tilde{H}^\ell &= Z^\ast_p \sigma^{\ell p}_{p'} Z^{p'} \,.
\end{align}
Note that all fields in the holon metal approach are neutral under U(1)$_1$ and U(1)$_2$, and only carry gauge charges under SU(2)$_S$ \cite{sdw09}.
%%%%%%%%%%%%%%%%%%%%%%%%%%%%%%%%%%%%%%%%%%%%%%%%%%%%%%%%%%%%%%%%%%

\bibliography{bibliography}
\end{document}